\documentclass[12pt,preprint]{aastex}

\slugcomment{Version: \today}

\shorttitle{NIRSPEC echelle observations of V1647~Orionis}
\shortauthors{ASPIN, GREENE \& REIPURTH}

\begin{document}

\title{V1647~ORIONIS: KECK/NIRSPEC 2~MICRON ECHELLE OBSERVATIONS}

\author{COLIN~ASPIN\altaffilmark{1}$^{,3}$, THOMAS P. GREENE\altaffilmark{2}$^{,3}$, BO REIPURTH$^1$}

\affil{$^1$ Institute for Astronomy, University of Hawaii,\\
  640 N. A'ohoku Place, Hilo, HI 96720 \\
  {\it caa@ifa.hawaii.edu, reipurth@ifa.hawaii.edu}}

\affil{$^2$ NASA Ames Research Center, MS 245-6, \\
  Moffett Field, CA 94035-1000 \\
{\it thomas.p.greene@nasa.gov}}

\altaffiltext{3}{Visiting Astronomers at the W.M.~Keck~Observatory.}

\begin{abstract} 
  We present new Keck~II NIRSPEC high-spectral resolution 2~$\mu$m echelle observations of the young eruptive   variable star V1647~Orionis.  This star went into outburst in late 2003 and faded to its pre-outburst brightness after approximately 26 months.  V1647~Orionis is the illuminating star of McNeil's Nebula and is located near M~78 in the Lynds 1630 dark cloud.  Our spectra have a resolving power of approximately 18,000 and allow us to study in detail the weak absorption features present on the strong near-IR veiled   continuum.  An analysis of the echelle orders containing Mg~I (2.1066~$\mu$m) and Al~I (2.1099~$\mu$m), Br$\gamma$ (2.1661~$\mu$m), the Na~I doublet (2.206 and 2.209~$\mu$m), and the CO overtone bandhead   (2.2935~$\mu$m) gives us considerable information on the physical and geometric characteristics of the regions producing these spectral features.  We find that, at high-spectral resolution, V1647~Orionis in quiescence resembles a significant number of FU~Orionis type eruptive variables and does not appear similar to the quiescent EX~Lupi variables observed.  This correspondence is discussed and implications   for the evolutionary state of the star are considered.
\end{abstract}


\section{INTRODUCTION}
The 2003--2006 outburst of the partly embedded, pre-main sequence star V1647~Orionis has provided a plethora of information on the nature of eruptive variables and the physical changes undergone by a star during such an event.   V1647~Ori, the illuminating source of McNeil's Nebula (McNeil 2004) located near M~78 in the Lynds 1630 dark cloud, brightened by over 5 magnitudes in the optical (Brice\~no et al. 2004; Reipurth \& Aspin 2004a) and 3 magnitudes in the near-IR (Reipurth \& Aspin 2004a; Ojha et al. 2004) in the space of a few weeks.  The eruption also produced an enhanced H~I spectrum highlighted by a high-velocity ($\sim$600~km~s$^{-1}$) blue-shifted H$\alpha$ absorption component creating a P~Cygni profile (Reipurth \& Aspin 2004a; Walter et al. 2004; Fedele et al. 2007).  In the near-IR, strong $v=2-0$ CO overtone bandhead emission was observed immediately after outburst (Reipurth \& Aspin 2004a; Vacca et al. 2004) which faded as the event progressed to weak absorption (Aspin, Beck, \& Reipurth 2008, henceforth ABR08; Brittain et al. 2007).   The pre-outburst, outburst, and post-outburst spectral energy distribution (SED) of V1647~Ori was studied by \'Abr\'aham et al. (2004), Andrews, Rothberg, \& Simon (2004), and ABR08, respectively.  Pre-outburst, the SED was that of a flat-spectrum source with an integrated bolometric luminosity (L$_{bol}$) of $\sim$5.6~L$_{\odot}$.  Soon after the outburst occurred, L$_{bol}$ had risen by a factor 5 to 10.  Approximately one year after the optical brightness of the star had returned to its pre-outburst level, V1647~Ori had an L$_{bol}\sim$9.2~L$_{\odot}$.  This outburst of V1647~Ori was the second documented event for this source; Aspin et al. (2006) studied a previous event, caught on photographic plates and film, from 1966--1967.  They concluded that both events were very similar in timescale and amplitude suggesting a common origin for the eruptions.  To date, 40+ refereed publications have been published on the 2003--2006 outburst covering all wavelengths from X-ray to optical to sub-mm and detailing the variability observed.   The reader is referred to the list of references in ABR08 for more detailed information on the specific waveband observations. 

Numerous authors have attempted to classify the outburst of V1647~Ori as either an FU~Orionis (FUor) or EX~Lupi (EXor) eruptive event.  These designations were first defined and discussed by Herbig (1977, 1989).  FUor events are deemed to be long-term (decades) eruptions, commonly commencing with a rapid brightening followed by a slow decline as observed in the class prototype, FU~Orionis.  EXor eruptions are seen to be shorter term events (typically lasting months) with similar brightening characteristics yet highly variable in decline as observed in the class prototype, EX~Lupi (Herbig 1977; Herbig et al. 2001).  Much discussion has transpired regarding these two types of outbursts, specifically whether they are the result of the same physical event acting on different timescales, or different triggering mechanisms producing different observed characteristics.  In fact, the mechanism by which FUor events are initiated is not yet definitively known, with several mechanisms being proposed.  Thermal runaway accretion in a circumstellar accretion disk was considered by Bell \& Lin (1994) and Bell et al. (1995), whereas the evolution from envelope accretion to (overloaded) disk magnetospheric accretion was proposed by Hartmann \& Kenyon (1985, 1996).  Alternatively, a mechanism involving the close approach of a stellar companion in an elliptical orbit which disrupts the inner regions of the accretion disk was described by Bonnell \& Bastien (1992) and elaborated on by Reipurth \& Aspin (2004b).  Finally, Gammie (1996) suggested that perhaps intra-disk gaps modulate accretion bursts.  Regarding V1647~Ori, the majority of authors have classified its outburst as an EXor rather than FUor event due to its short duration and observed characteristics.  However, it is perhaps not clear that two distinct categories of eruptive events exist, rather a {\it continuum} of events may occur with the FUors being at one extreme (longest) and the EXors at the other (shortest).  Gibb et al. (2006) and Fedele et al. (2007) also suggested such a scenario.  Additionally, perhaps factors such as source age and disk mass may play a role in determining the nature of outbursts. 

In this paper we will show that V1647~Ori possesses high-spectral resolution near-IR characteristics almost identical to a number of known FUors and FUor-like objects (objects in which the eruption was unobserved).  Due to the strong near-IR (veiled) continuum present in the spectrum of V1647~Ori, and the weak nature of, for example, gravity- and temperature-sensitive absorption features (ABR08), high-spectral resolution observations are not only advantageous but a necessity.   In $\S 2$ we present our new near-IR (R$\sim$18,000) observations made using the Keck~II 10-meter telescope and NIRSPEC.  In $\S 3$ we compare these observations to those of other sources specifically a sample of FUors, quiescent EXors, and MK standard stars.  In $\S 4$ we discuss our findings in relation to the FUor/EXor classification scheme and draw conclusions on the nature of the V1647~Ori outbursts.

\section{OBSERVATIONS AND DATA REDUCTION}
We have used the Keck~II 10-meter telescope on Mauna Kea, Hawaii with the facility near-IR multi-order cryogenic spectrograph NIRSPEC (McLean et al. 1998).  Observations of V1647~Ori were obtained on UT August 26, 2007 and January 26, 2008 via remote operation from the Waimea Keck~II control room.  Table~\ref{jobs} presents the complete observation log and includes data taken on previous runs and used for comparative purposes.  Spectra were acquired using a 4-pixel (0$\farcs$576) wide slit resulting in a pixel-defined spectral resolution, R, of $\sim$18,000 or, at 2~$\mu$m, 16.7~km~s$^{-1}$).  The pixel scale was 0$\farcs$2~pixel$^{-1}$ along the 12$''$ longslit and astronomical seeing was $\sim$0$\farcs$6 during the observations.  The NIRSPEC grating was positioned to an echelle angle of 61.89$^{\circ}$ and a cross-disperser angle of 35.45$^{\circ}$ resulting in echelle orders 32--38 falling on the science array.  These included the spectral regimes containing Mg~I (2.1066~$\mu$m, order~\#36), Al~I (2.1099~$\mu$m, order~\#36), H~I Br$\gamma$ (2.1661~$\mu$m, order~\#35), Na~I (2.206 and 2.209~$\mu$m, order~\#34), and the v=2-0 CO overtone bandhead (2.2935~$\mu$m, order~\#33).  The NIRSPEC-7 blocking filter was also used. The free spectral range of each echelle order was $\Delta\lambda\simeq\lambda$/67 or $\Delta$v$\sim$4450~km~s$^{-1}$.

During the on-sky exposures, the slit was physically stationary and hence was allowed to rotate on the sky during alt-az tracking.  Data were acquired in ABBA sequences with the target on the slit at all times and nodded by $\pm$6$''$.  The total exposure time on V1647~Ori (m$_K\sim$10) was 20~minutes. Observations of the A0~V star HIP~29881 (m$_K\sim$6.6) were obtained immediately following the target observations and at a similar airmass to use as a ratio star to remove telluric features.  The standard ``Lamps ON'' (flat, dark, arc) calibration sequence was obtained at the start of the observing night and used for flat-fielding and wavelength calibration.  Telluric OH emission lines were additionally used to obtain an accurate wavelength calibration.  We utilized the telescope autoguider fed by the internal IR camera ``SCAM'' to keep the targets on the long-slit.

Data reduction was performed using the Starlink {\tt FIGARO} spectroscopic reduction package (Shortridge et al. 2002).  First, AB pairs were subtracted, and the resultant images flat-fielded.  Next, the programs {\tt profile} and {\tt optextract} were used to optimally extract the source spectra.  Wavelength calibration was performed using the program {\tt arc} and then applied to the extracted spectra using the program {\tt xcopy}. The resultant spectra were then ratioed with the extracted standard star spectrum using program {\tt irflux} and the flux in Janskys was converted to W~m$^{-2}$~$\mu$m$^{-1}$ using program {\tt irconv}. The Br$\gamma$ absorption feature in the A0~V standard was removed before application to the target spectra using linear interpolation and program {\tt isedit}.  Due to the width of the Br$\gamma$ absorption in the standard, some weak telluric absorption features remained in the ratioed target spectrum near 2.1661~$\mu$m.  However, since they do not affect the profile of the Br$\gamma$ emission in V1647~Ori and the Br$\gamma$ line is corrected for absorption in the telluric standard, further consideration of these features was not undertaken.   Finally, the four (ABBA) ratioed, flux calibrated spectra of V1647~Ori were median filtered to remove noise spikes.  This improved the signal to noise in the final spectrum by approximately a factor $\times$2. 

We have corrected the final spectra for the {\it v}$_{helio}$ radial velocity component using the values obtained from the iraf tasks {\tt rvcorrect} with the UT time of observation to determine the value and {\tt specshift} to shift the spectra in wavelength by the appropriate amount.

Below, we compare and contrast the V1647~Ori echelle spectral orders to the equivalent data on several classical FUors, FUor-like stars, EXors, one other young Class~I protostellar object, and MK standard stars.  The sample of comparison sources discussed here were selected from a high resolution spectral survey of FUors and EXors undertaken by the authors.  Additional data is used from previous observing runs by one of us (TRG).  All the data used (with the exception of the $\alpha$~Ori spectrum, see below for the origin of these data) was taken using NIRSPEC in the identical configuration as described above.  Of these stars, the classical FUors FU~Ori, L1551~IRS5, V883~Ori, and V1057~Cyg are well-known and well studied examples of the group. The FUor-like objects HH354~IRS, HH381~IRS, and Par~21, have many of the characteristics of classical FUors but were not observed in their brightening phase.  The above FUors and FUor-like objects are further described in Reipurth \& Aspin (1997).  The stars V1118~Ori, V1143~Ori, NY~Ori, and VY~Tau are short-term eruptive variables of the EXor group (Herbig, private communication).  The paper by Herbig (2008) discusses most of these EXors in detail.  The Class~I source that we have included as a comparison object is $\rho$~Oph IRS63 which was observed and analyzed by Doppmann et al. (2005).  The MK standards used are described below.

\section{RESULTS}

\begin{figure*}[tb] 
\epsscale{0.8}
\plotone{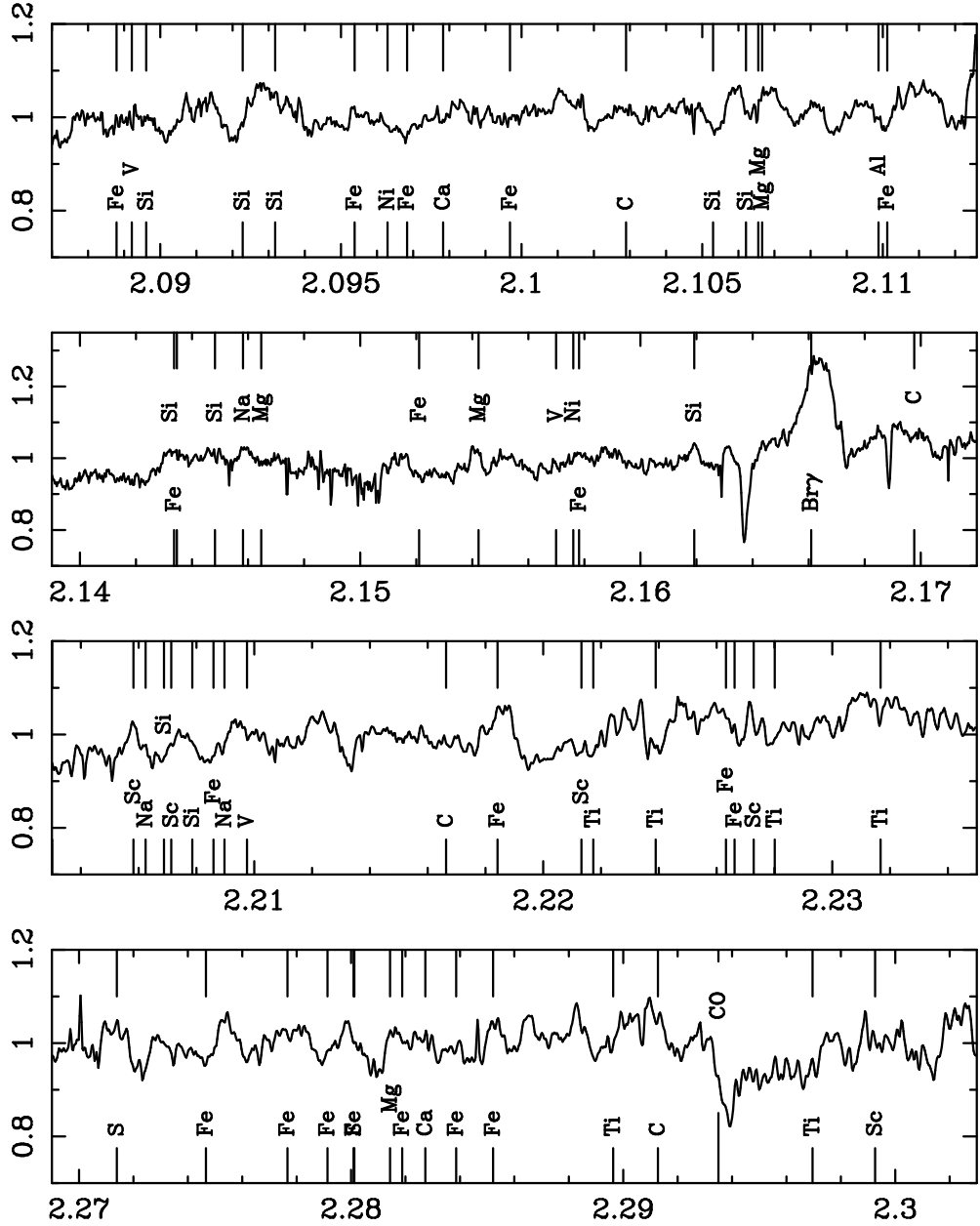} 
\caption{The V1647~Ori NIRSPEC echelle orders 36, 35, 34, and 33 showing Mg+Al, (top), Br$\gamma$ (upper middle), Na~I (lower middle), and CO (bottom), respectively.  The spectra have been corrected for the $v_{helio}$ radial velocity at the time of observation.  All atomic features plus the molecular CO bandhead are indicated.  The two sharp absorption features on either side of the Br$\gamma$ emission line are telluric absorption lines that were not removed by ratioing due to the interpolation over the A0~V standard star spectrum B$\gamma$ absorption feature.  These telluric features serve to demonstrate the full-width half maximum of unresolved lines in relation to the broad features seen in the V1647~Ori spectrum. 
\label{orders4}}

\end{figure*}

Of the seven echelle orders observed using NIRSPEC in the above configuration (32--38), we shall consider four which contain important spectral features related to the physics of the region in which they are produced.  These are orders \#36 containing Mg~I (2.1066~$\mu$m) and Al~I (2.1099~$\mu$m), \#35 containing Br$\gamma$ (2.1661~$\mu$m), \#34 containing the Na~I doublet (2.206 and 2.209~$\mu$m), and \#33 containing the $v=2-0$ CO overtone bandhead (2.2935~$\mu$m).  The final spectra of these four orders for V1647~Ori (2007 Aug 26) are shown in Fig.~\ref{orders4}.  The top panel shows order \#36, the upper-middle panel order \#35, the lower-middle panel order \#34, and the bottom panel order \#33 and all atomic spectral features present are indicated.  Of particular note in these echelle orders is the fact that all spectral features appear broad with respect to the residual sky absorption line at $\sim$2.1638~$\mu$m (its presence is explained below).  We estimate that the unresolved sky absorption line has a full-width half maximum (FWHM) of 20~km~s$^{-1}$ while, for example, the Br$\gamma$ emission line has a FWHM of 150~km~s$^{-1}$.  We additionally note that the $v=2-0$ CO overtone bandhead at 2.2935~$\mu$m also appears significantly broadened.  

\subsection{Echelle Order \#34 -- the Na D Doublet}
\begin{figure*}[tb] 
\epsscale{1.0}
\plotone{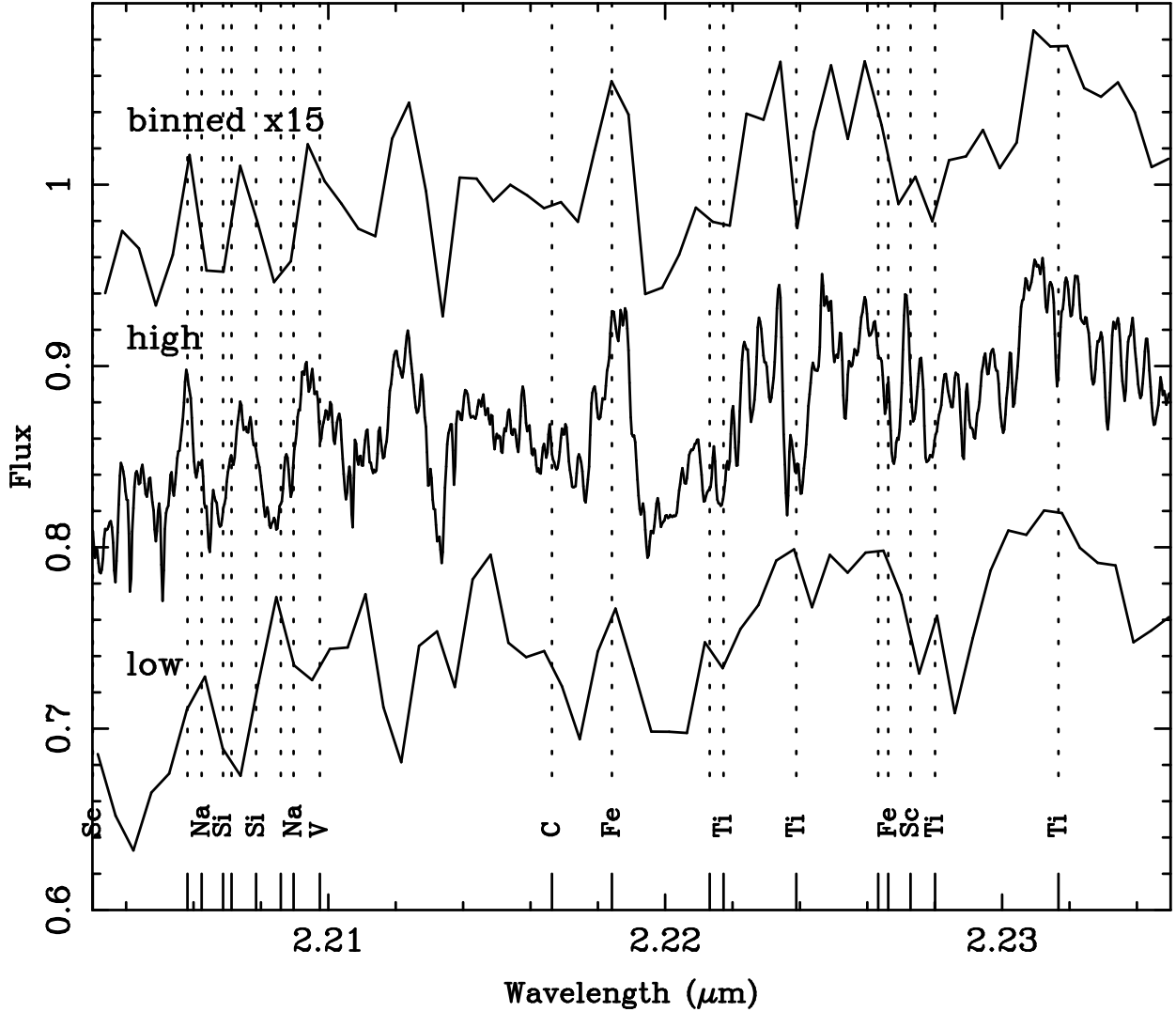} 
\caption{A comparison of the NIRSPEC echelle R$\sim$18,000 Na order with the low-resolution NASA IRTF SpeX data used in ABR08 to investigate the physics of the absorbing region in V1647~Ori.  The spectra have been corrected for the $v_{helio}$ radial velocity at the time of observation.  We show both the full-resolution NIRSPEC spectrum (middle), that spectrum binned (by a factor $\times$15) to approximately match the resolution of the IRTF data (top), and the IRTF spectrum (bottom).  Note that there is general agreement in the shape of the two spectra taken 5 months apart although there appears to be a small (yet significant when compared to the R$\sim$18,000 data) dispersion change across the low-resolution IRTF spectrum. 
\label{hi-lores}}

\end{figure*}

In Fig.~\ref{hi-lores} we show the NIRSPEC Na echelle order together with the same wavelength range from our low-resolution NASA IRTF SpeX spectrum of V1647~Ori from February 2007 and presented in ABR08.  This clearly demonstrates the superior ability of high-resolution spectroscopy in the study of weak absorption features on a strong continuum.  The features observed at high-resolution are, in general, mirrored in the low-resolution spectrum although some differences are present.  This is perhaps not unexpected since the observations were taken 5 months apart but could be additionally due to small wavelength calibration errors in the low-resolution spectrum resulting in a slightly variable dispersion.  We show this comparison since our low-resolution spectrum was used (see ABR08) to derive effective temperature (T$_{eff}$) and spectral type (M0~V) for the absorbing region which was assumed to be the stellar photosphere of V1647~Ori.

\begin{figure*}[tb] 
\epsscale{0.75}
\plotone{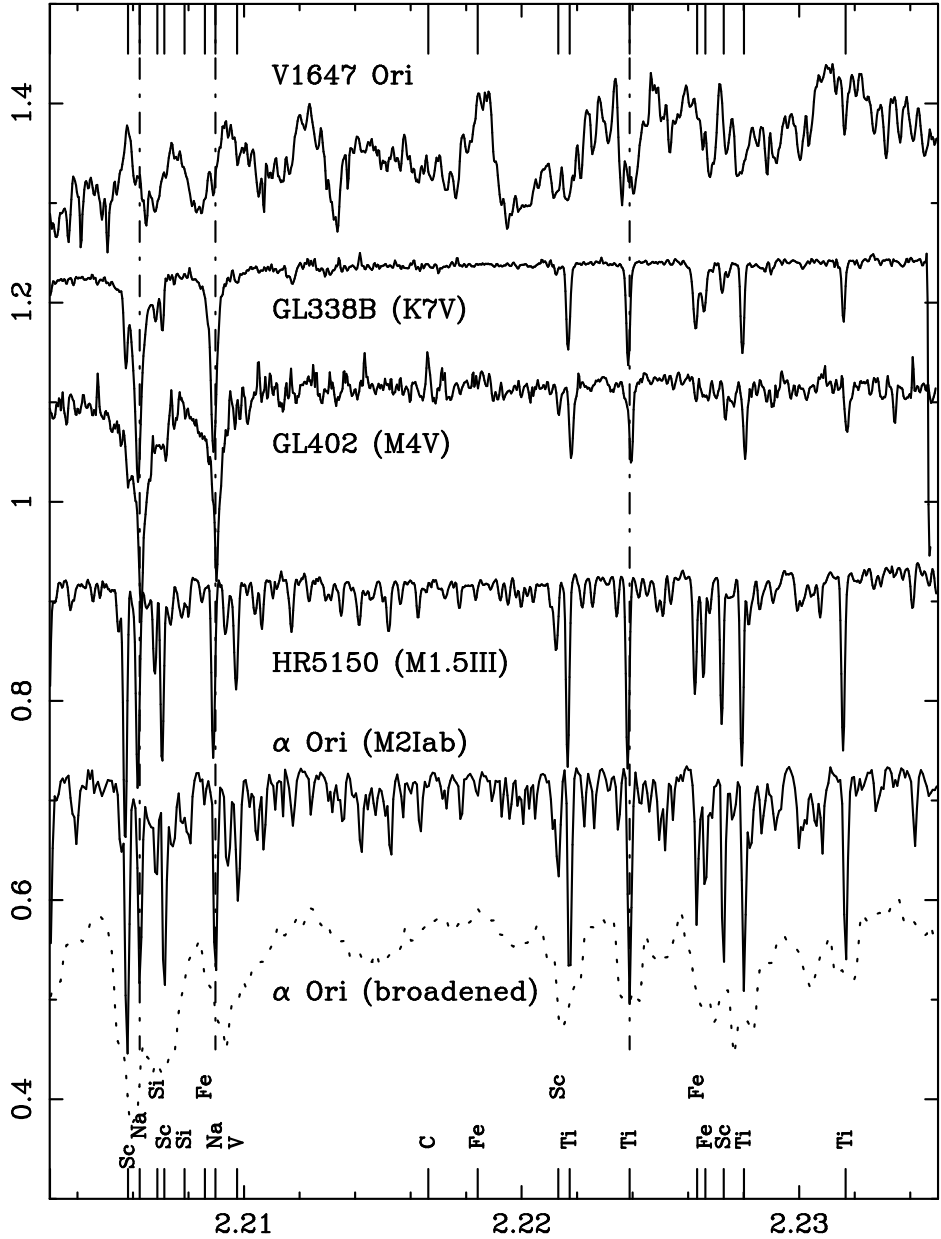} 
\caption{The V1647~Ori Na echelle order (top) together with the same wavelength range from MK standard stars GL~338B (K7~V), GL~402 (M4~V), HR~5150 (M1.5~III), and $\alpha$~Ori (M2~Iab).  The spectra have been corrected for the $v_{helio}$ radial velocity at the time of observation.  Although the features seen in the standards are mostly present in V1647~Ori, they are considerably sharper in the MK standards suggesting that they are considerably broadened in V1647~Ori.  The bottom dotted line is the $\alpha$~Ori spectrum artificially broadened to simulate the V1647~Ori data.  The broadening was performed by smoothing the $\alpha$~Ori spectrum using the IRAF task {\tt splot} and the 's' command.  The smoothing used was $\sim$120~km~s$^{-1}$ and gave a good match between the isolated Ti line at 2.224~$\mu$m in both sources.  All atomic lines in this wavelength regime are identified at the bottom of the plot.
\label{na-mk}}

\end{figure*}

It is clear from Fig.~\ref{na-mk} that, compared to MK spectral standard stars, specifically the K7~V dwarf GL~388B, the M4~V dwarf GL~402, the M1.5~III HR~5150, and the M2~Iab supergiant $\alpha$~Ori, the Na order of V1647~Ori is fundamentally different in its spectral characteristics.  The data on GL~338B, GL~402, and HR~5150 were taken from Doppmann et al. (2005) who utilized the same NIRSPEC setup as our V1647~Ori observations.  The $\alpha$~Ori data are from Wallace \& Hinkle (1996) and were taken at a factor $\sim$4 higher spectral resolution than our NIRSPEC observations.   However, they were degraded to match the spectral resolution of our data prior to use.  Whereas the MK standards show relatively narrow (almost unresolved) atomic absorption features with FWHM of $\sim$20~km~s$^{-1}$ (the Na lines are pressure broadened in the dwarf spectrum), the features in V1647~Ori are very broad and blended with a FWHM of typically $\sim$120~km~s$^{-1}$.   The significant structure seen in V1647~Ori has little in common with the observed dwarf, giant, or supergiant star spectra and is perhaps more reminiscent of a significantly broadened giant/supergiant which contain a wealth of molecular lines from their low-surface gravity atmospheres.  To investigate this, we have artificially broadened the $\alpha$~Ori spectrum using the smooth feature ('s') in the IRAF task {\tt splot}.  The dotted line at the bottom of the plot is the $\alpha$~Ori spectrum smoothed to give a Ti 2.224~$\mu$m line width of $\sim$120~km~s$^{-1}$.  As one can see, although the features do not all correlate well with the V1647~Ori spectrum, the general trend and gross shape of the spectra are similar.  

\begin{figure*}[tb] 
\epsscale{0.8}
\plotone{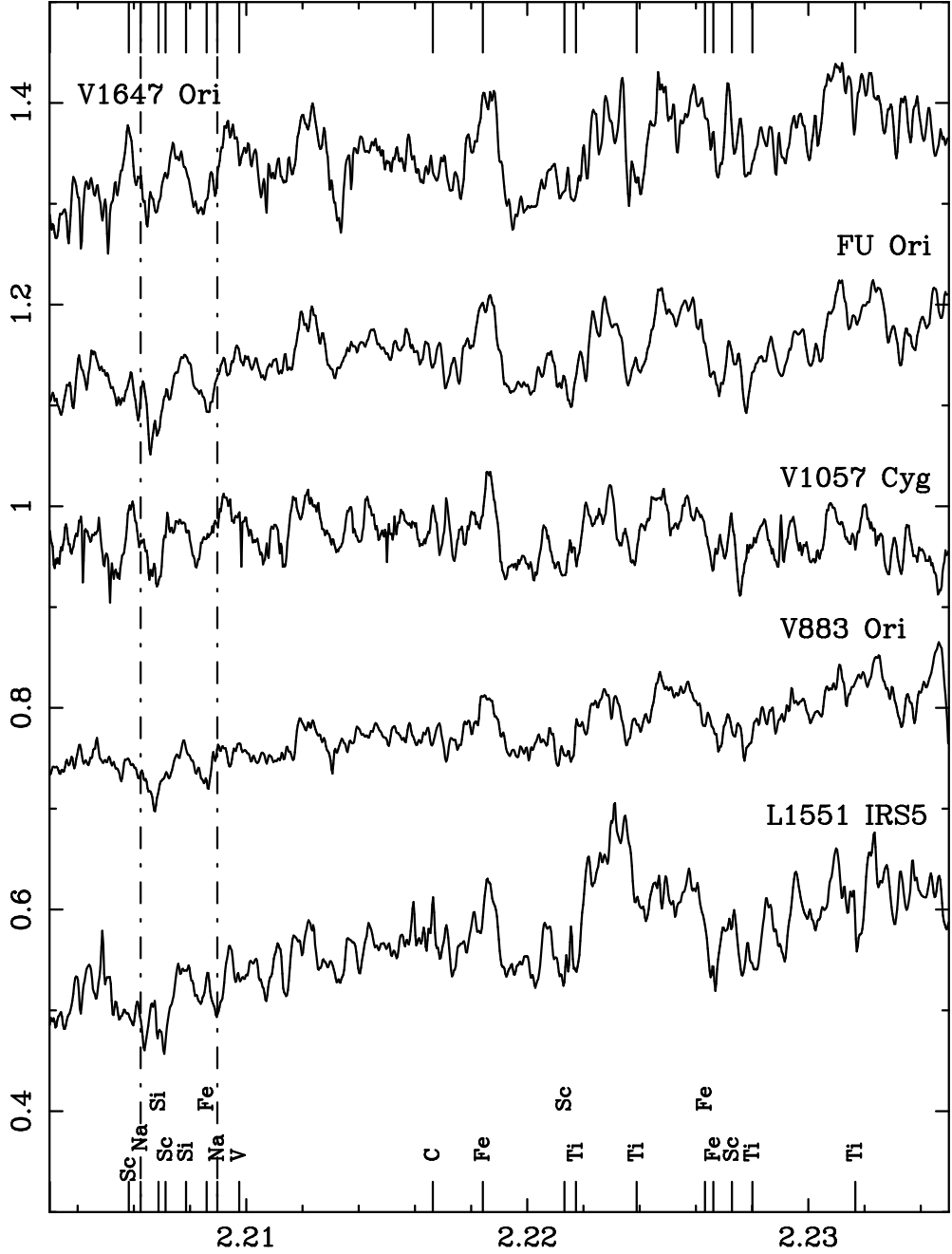} 
\caption{Comparison of the Na echelle order spectrum from V1647~Ori with the same wavelength regime from two known FUors, FU~Ori itself and V1057~Cyg, and two FUor-like objects observed by Reipurth \& Aspin (1997) and confirmed as FUors by GAR08.  These objects are V883~Ori and L1551~IRS5.  The spectra have been corrected for the $v_{helio}$ radial velocity at the time of observation. The comparison between the features present in these objects is quite remarkable and suggests that V1647~Ori in quiescence and the active FUors have much in common.  Again, all atomic features in this wavelength range are identified and additionally the Na doublet at 2.206 and 2.208~$\mu$m are marked with dot-dashed lines through all spectra as an aid to relating features between them. 
\label{na-fuors-rvcor}}

\end{figure*}

In Fig.~\ref{na-fuors-rvcor} we have plotted the Na order spectrum of V1647~Ori with similar spectra of the classical FUors FU~Ori, V1057~Cyg, V883~Ori, and L1551~IRS5 all taken from Greene, Aspin, \& Reipurth (2008, henceforth GAR08) with the exception of V1057~Cyg which was observed on the same night as our first V1647~Ori spectrum.  Using the iraf rv package task {\tt rvcorrect}, we have corrected the spectra for the appropriate {\it v}$_{helio}$ radial velocity component.\footnotemark\footnotetext{However, we have not corrected for the difference in molecular cloud radial velocities. } The most striking feature of this plot is the fact that {\it all} spectra appear remarkably similar, in particular the echelle spectrum of FU~Ori itself appears almost identical to that of V1647~Ori.  To quantify this similarity, we have performed a cross-correlation (henceforth referred to as xc) analysis of these spectra (and those of the MK standards above) using the IRAF rv package task {\tt fxcor} and the results are presented in Table~\ref{xcorrnaco}.  The xc value used to determine the correlation between spectra is the Tonry \& Davis (1979) R value which gives a measure of the correlation peak height with respect to the average noise in the spectrum for each correlation function derived.  The R value is an output parameter of the {\tt fxcor} task.  We note that an R value $>$3 indicates a significant correlation between spectra and is approximately equivalent to a significance of $\sim$3$\sigma$.  Therefore, the larger the R value, the better the correlation.  We have also included the FUor-like objects HH354~IRS, HH381~IRS, and Par~21 (also taken from GAR08) in the xc analysis.  For the sake of brevity, we henceforth refer to both classical FUors and FUor-like objects by the generic name FUors. 

\begin{figure*}[tb] 
\epsscale{0.9}
\plotone{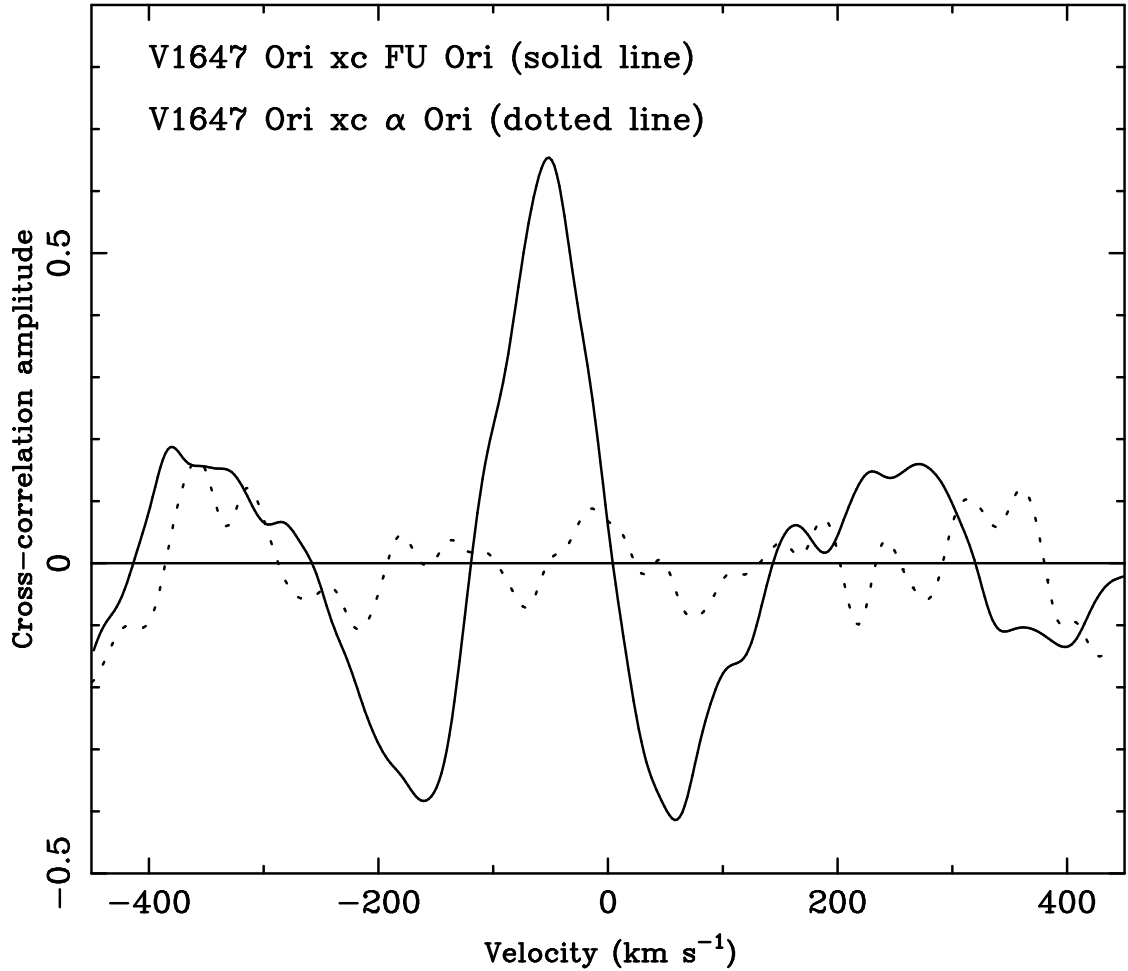} 
\caption{A cross-correlation function from the comparison of the V1647~Ori Na echelle order with that of FU~Ori (solid line).  The sharp, symmetric peak at slightly negative velocity indicates an excellent correlation between these two sources and produces a Tonry \& Davis (1979) R value of 11.2.  Values above R=3 indicates a significant correlation, the higher the value, the better the correlation.  Also shown is the correlation function for the comparison of V1647~Ori and the observed $\alpha$~Ori data (dotted line).  Note the lack of any strong peak, indicating a generally poor correlation with R=0.6.
\label{xcorr}}

\end{figure*}

In echelle order \#34, the MK standard stars give very low R-values (R$<$1.1) when cross-correlated with V1647~Ori.  Additionally, the broadened $\alpha$~Ori spectrum does not correlate well with V1647~Ori producing R$<$1.  On the other hand, all the FUors shown in Fig.~\ref{na-fuors-rvcor} produce highly significant correlation values, with all five sources producing R values in the range 5.6 to 20.9.  Three FUors (FU~Ori, HH381~IRS, and Par~21) give R$>$12.  The mean R value over the seven FUor observations is $<$R$>$=11.3$\pm$5.2.  Clearly, the uncertainty quoted on $<$R$>$ indicated the range of R values encountered.   As a representative plot, we show the xc function for FU~Ori (solid line) and $\alpha$~Ori (dotted line) in Fig.~\ref{xcorr}.  We used the observed spectra (prior to $v_{helio}$ correction) in the xc analyses, and hence the velocity shift of the xc peak (for FU~Ori) is due to the sum of the difference in {\it v$_{helio}$} between sources at the times of their observation, and the difference in the velocity of the clouds in which they reside.   

\begin{figure*}[tb] 
\epsscale{0.8}
\plotone{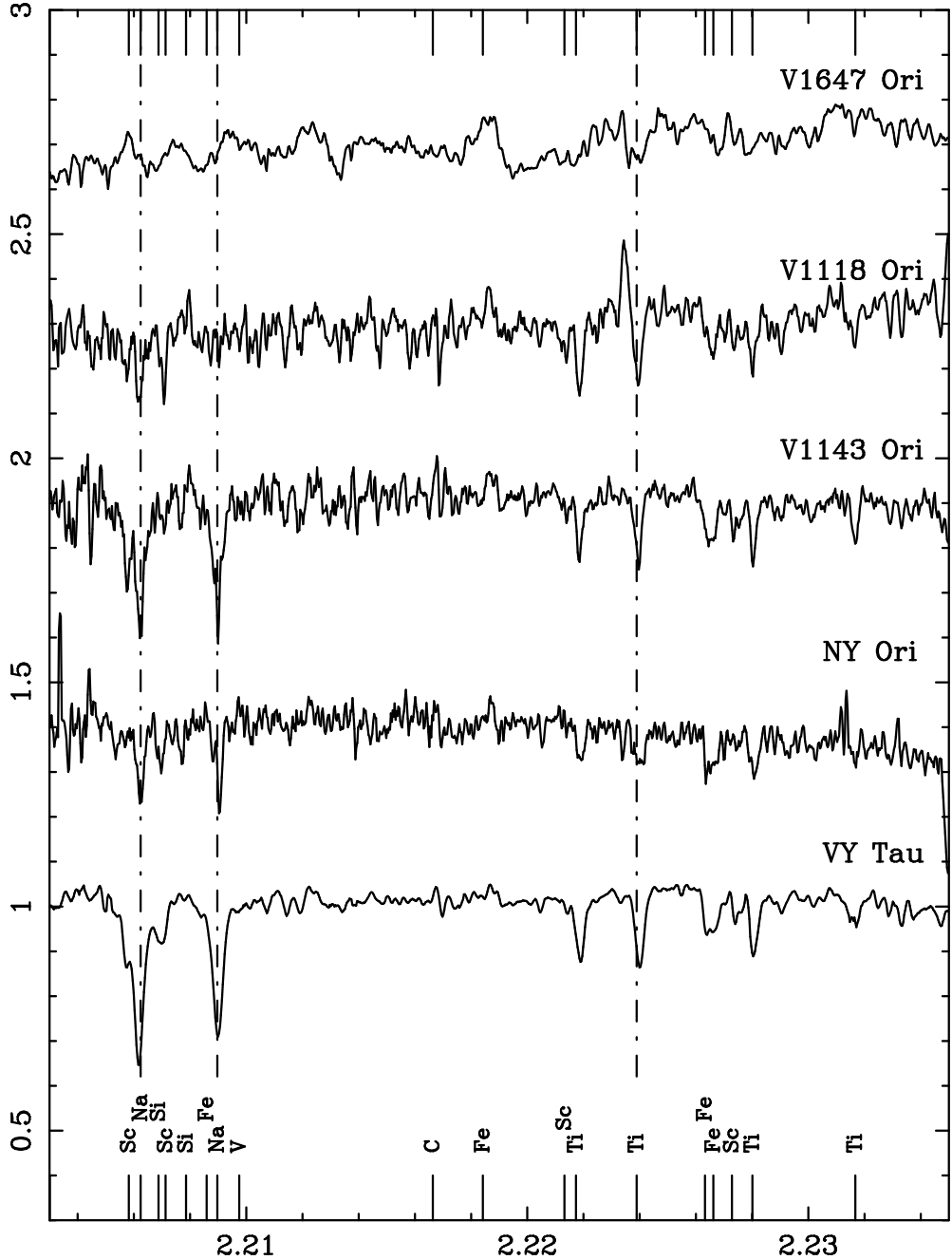} 
\caption{The spectra of the four EXor variables V1118~Ori, V1143~Ori, NY~Ori, and VY~Tau together with that of V1647~Ori.  In this plot we show the echelle order containing the Na~I lines at 2.206 and 2.209~$\mu$m.  The spectra have been corrected for the $v_{helio}$ radial velocity at the time of observation.  The vertical dot-dashed lines show the location of the two Na~I lines and a Ti~I line at 2.2249~$\mu$m and are provided only to guide the eye between the different spectra.  In all cases, the spectral features present in the EXors are considerably narrower than in V1647~Ori and, whereas the EXors absorption lines can be readily identified, it is much more difficult for V1647~Ori.  
\label{na-exors}}

\end{figure*}

In addition to the FUors, we have cross-correlated the echelle spectra of V1647~Ori with a sample of EXor variables and one heavily veiled (r$_K$=1.7) K5--7 Class~I protostar, $\rho$~Oph IRS~63. The latter source was found to have a rotational velocity of {\it v}~sin~{\it i}~$\sim$45~km~s$^{-1}$ by Doppmann et al. (2005). Fig.~\ref{na-exors} shows the Na order for V1647~Ori and four EXors, V1118~Ori, V1143~Ori, NY~Ori and VY~Tau.  All were observed while they were faint i.e. not in an eruptive state.  The values of R from the xc of these sources with V1647~Ori are in the range 0.8 to 3.9 with a mean value of $<$R$>$=1.8$\pm$1.2.  Only one source produced an R$>$3 correlation and therefore a statistically significant result (V1118~Ori, R=3.9).  The value of $<$R$>$ for the EXors is therefore over 6$\times$ smaller than the corresponding value for the FUors.  A visual comparison of the V1647~Ori and EXor spectra shows that although some of the EXor features appear present in V1647~Ori, they are considerably narrower in the EXors (e.g. Ti~I at $\sim$2.24~$\mu$m).  Artificially broadening the EXor spectra to match the FWHM of typical features in V1647~Ori also results in relatively poor xc values.  We also see that the V1647~Ori spectrum possesses considerably more structure than those of the EXors.  This is perhaps attributable to the presence of broad molecular bands in V1647~Ori that are not observed in the EXor spectra.  We finally note that the xc of V1647~Ori with IRS~63 gave R$\sim$0.6.

\subsection{Echelle Order \#33 -- the CO Overtone Bandhead}
\begin{figure*}[tb] 
\epsscale{0.8}
\plotone{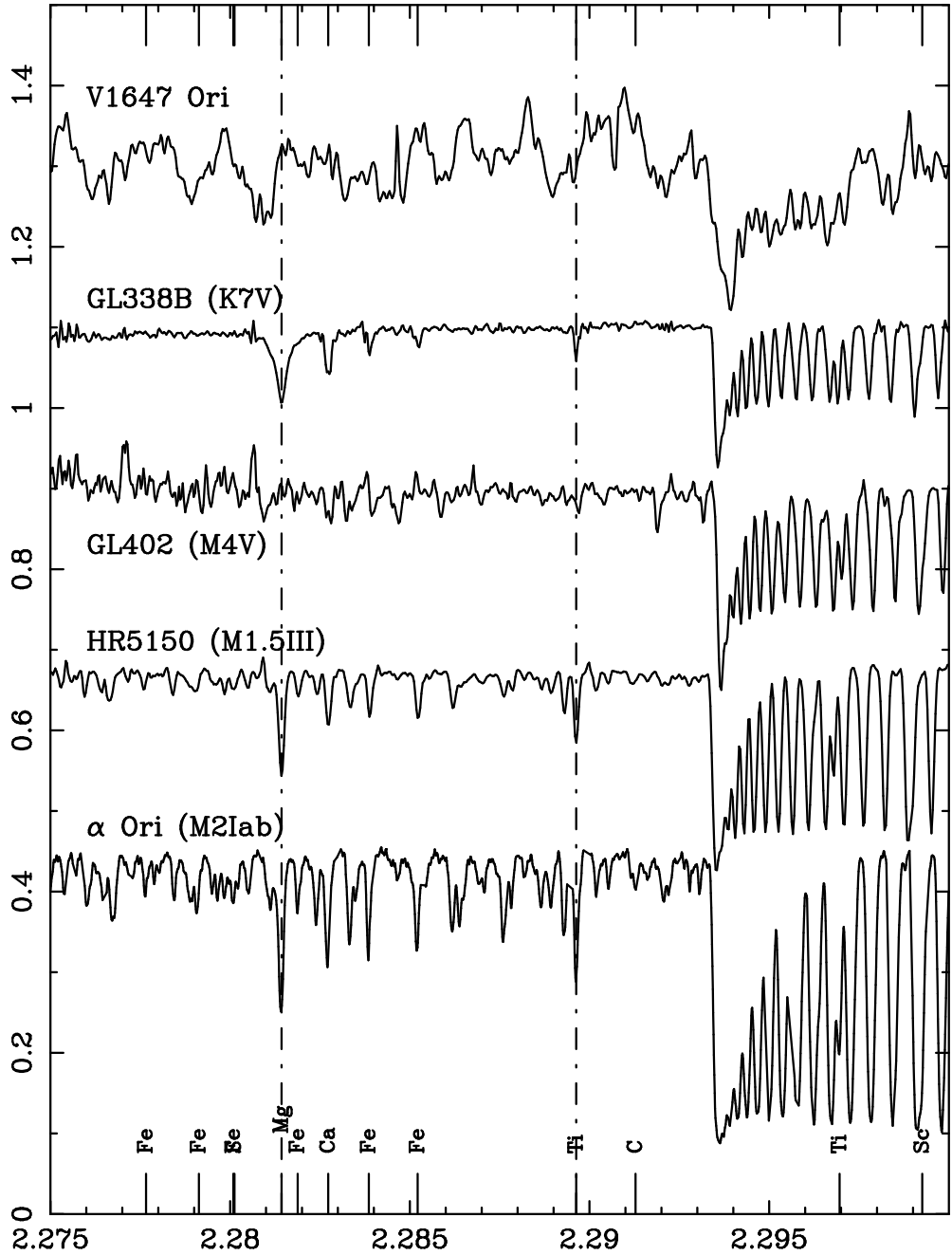} 
\caption{The same plot as in Fig.~\ref{na-mk} but for the CO bandhead echelle order.  The spectra have been corrected for the $v_{helio}$ radial velocity at the time of observation.  The same trends are evident as in the Na order, namely that the lines/bands in V1647~Ori are considerably broadened with respect to the corresponding lines/bands in the MK standards stars.  
\label{co-mk}}

\end{figure*}

A similar comparison can be performed using the echelle order containing the CO overtone bandhead.  Fig.~\ref{co-mk} shows the V1647~Ori observation together with similar observations of the MK standard stars.  Again, we have corrected the spectra for the appropriate {\it v}$_{helio}$ radial velocity components.  The features seen in the MK standard stars are considerably narrower than those in V1647~Ori and the CO bandhead itself is very broad in V1647~Ori with respect to its profile in the dwarfs, giants, and supergiants.  Support for this comes from the comparison of both the slope of the bandheads, and the fact that the individual CO lines longward of the bandhead are blended in V1647~Ori unlike in the MK standards.  

\begin{figure*}[tb] 
\epsscale{0.8}
\plotone{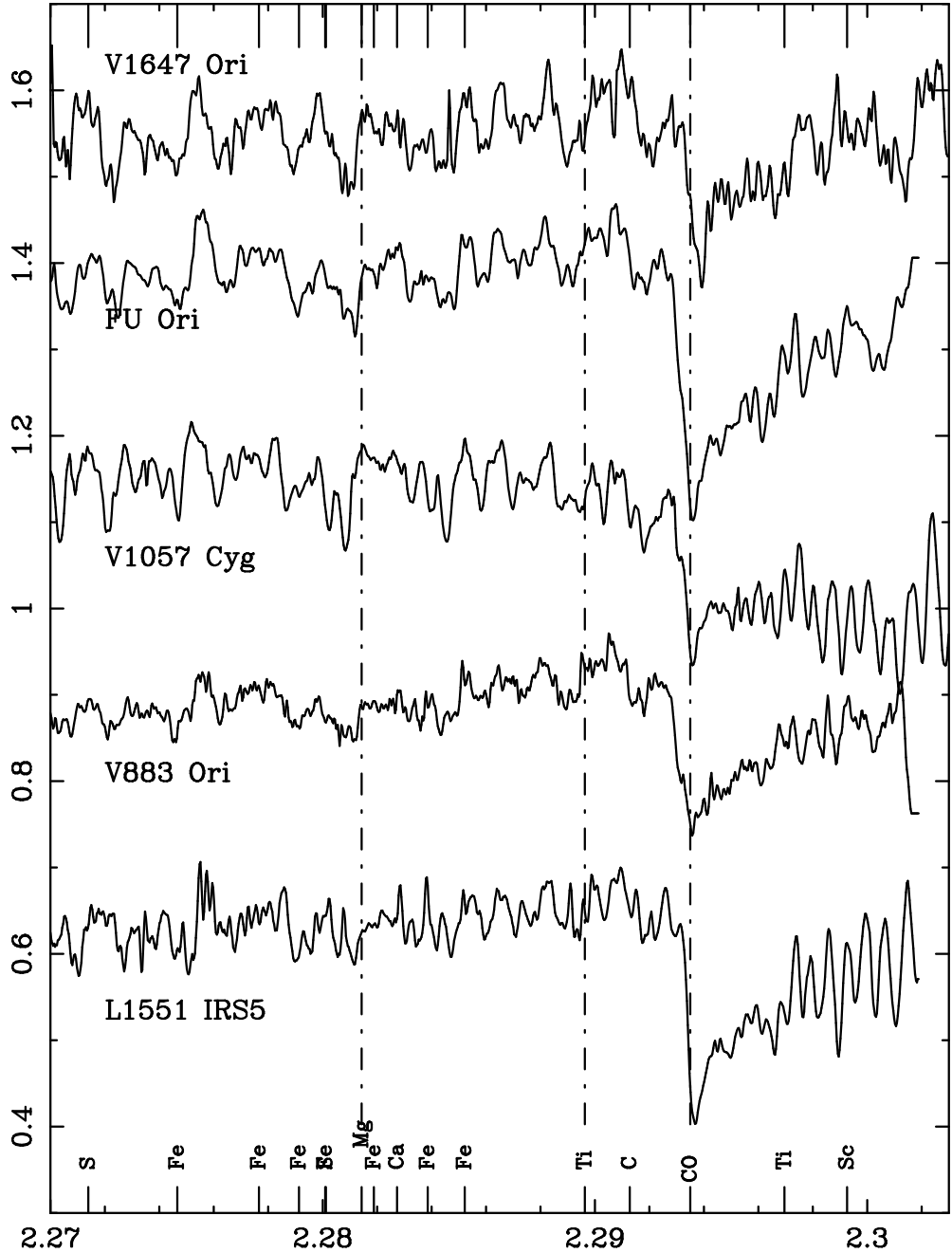} 
\caption{The same plot as in Fig.~\ref{na-fuors-rvcor} but for the CO bandhead echelle order.  The spectra have been corrected for the $v_{helio}$ radial velocity at the time of observation.  Note that as in the Na order, all features are broadened in all sources and V1647~Ori bears a striking resemblance to the FUors FU~Ori and V1057~Cyg, and the FUor-like objects V883~Ori and L1551~IRS5.  The dot-dashed lines show the location of the Mg line at 2.28143$\mu$m, the Ti line at 2.28963$\mu$m, and the CO bandhead at 2.2935~$\mu$m and are shown as guides to the eye through all the spectra.
\label{co-fuors}}

\end{figure*}

A comparison of the CO echelle order of V1647~Ori and the same FUors shown in Fig.~\ref{na-fuors-rvcor} is presented in Fig.~\ref{co-fuors}.  As in the Na region comparison, there is considerable correspondence between the features seen in all spectra although there is some obvious differences in line/band broadening, for example, L1551~IRS5 seems less broadened than either V1647~Ori, FU~Ori or V883~Ori.  The atomic features in this wavelength range are indicated at the bottom of the plot.  However, as with the Na echelle order, there are numerous features that do not correspond to atomic lines even if they are significantly broadened.  The spectra of $\alpha$~Ori shown in Figs.~\ref{na-mk} and \ref{co-mk} have many weak, narrow features.  The plot of $\alpha$~Ori from Wallace \& Hinkle (1996) together with their Table~2 suggests that the majority of these weaker features correspond to molecular CN.  In fact, 600 CN lines are present in the 2 to 2.4~$\mu$m passband together with 627 from CO ro-vibrational overtone transitions (longward of 2.2935~$\mu$m).  Perhaps molecular CN lines are also present in the V1647~Ori spectrum (as CO lines are).

\begin{figure*}[tb] 
\epsscale{0.8}
\plotone{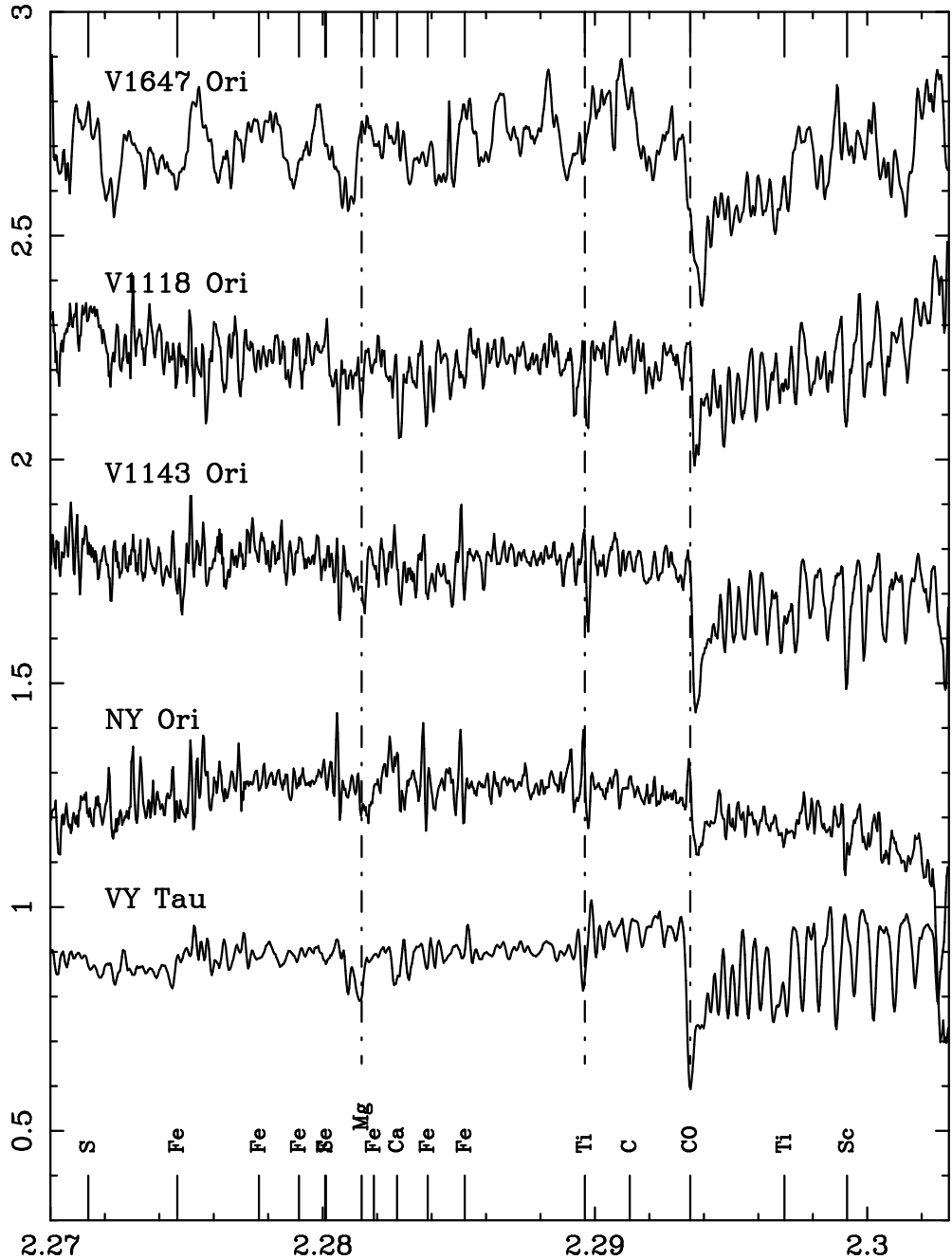} 
\caption{The same as Fig.\ref{na-exors} but for the echelle order containing the CO overtone bandhead at 2.2935~$\mu$m.  The spectra have been corrected for the $v_{helio}$ radial velocity at the time of observation.  Again, the dot-dashed lines identify several absorption lines and are provided to guide the eye between the different spectra.  Here, as in Fig.~\ref{na-exors}, the features in V1647~Ori are considerably broader than in the EXors.  Also, the CO bandhead is clearly sharper in the EXors than in V1647~Ori.
\label{co-exors}}

\end{figure*}

Quantitatively, we have cross-correlated the V1647~Ori CO spectrum with the MK dwarfs, giants, and supergiants, and all other objects from Table~\ref{xcorrnaco}. We have performed this xc over two wavelength ranges, first over the whole echelle order (including the CO bandhead), and second over a restricted wavelength range, specifically, 2.27 to 2.292~$\mu$m (excluding the CO bandhead).  The results of the restricted range xc are shown in parentheses immediately following the xc values for the whole echelle order.  Since the CO absorption bands are strong and extensive with respect to the atomic lines, we expect that in all objects that possess CO absorption, a significantly higher correlation value will be found when CO is included in the xc.  This is borne out by the xc values produced for the MK standards where, over the whole echelle order, a significant correlation of $<$R$>$=9.4$\pm$1.8 is found.   However, for the region shortward of the bandhead, the xc analysis indicates that V1647~Ori correlates less well (although still statistically significant) with the MK standards ($<$R$>=$3.3$\pm$0.7).   Further, as found with the Na order, V1647~Ori correlates very well with all FUors producing $<$R$>$=16.2$\pm$2.9 and 12.9$\pm$2.1 for the whole echelle order and the restricted spectral range, respectively.   Fig.~\ref{co-exors} shows the CO order spectra of V1647~Ori and the four EXors.  There is clearly more broad structure in the V1647~Ori spectrum than in the EXors and the CO bandhead itself is significantly broader in V1647~Ori.  The R values produced by the xc of V1647~Ori with the EXors gives $<$R$>$=9.2$\pm$4.0 and 4.1$\pm$1.0 for the full and restricted spectral region, respectively.  Again, as we found for the Na order spectra, the $<$R$>$ value obtained from the xc of V1647~Ori with the FUors is significantly larger (3.1$\times$) than the mean value obtained for the EXors (over the restricted spectral range not including the CO bandhead).  Finally we note that the xc of V1647~Ori with IRS~63 gives a values of R=13.6 and 5.0, for the full and restricted spectral regions, respectively.

\subsection{Echelle Order \#36 -- the Mg~I and Al~I Lines}
\begin{figure*}[tb] 
\epsscale{0.8}
\plotone{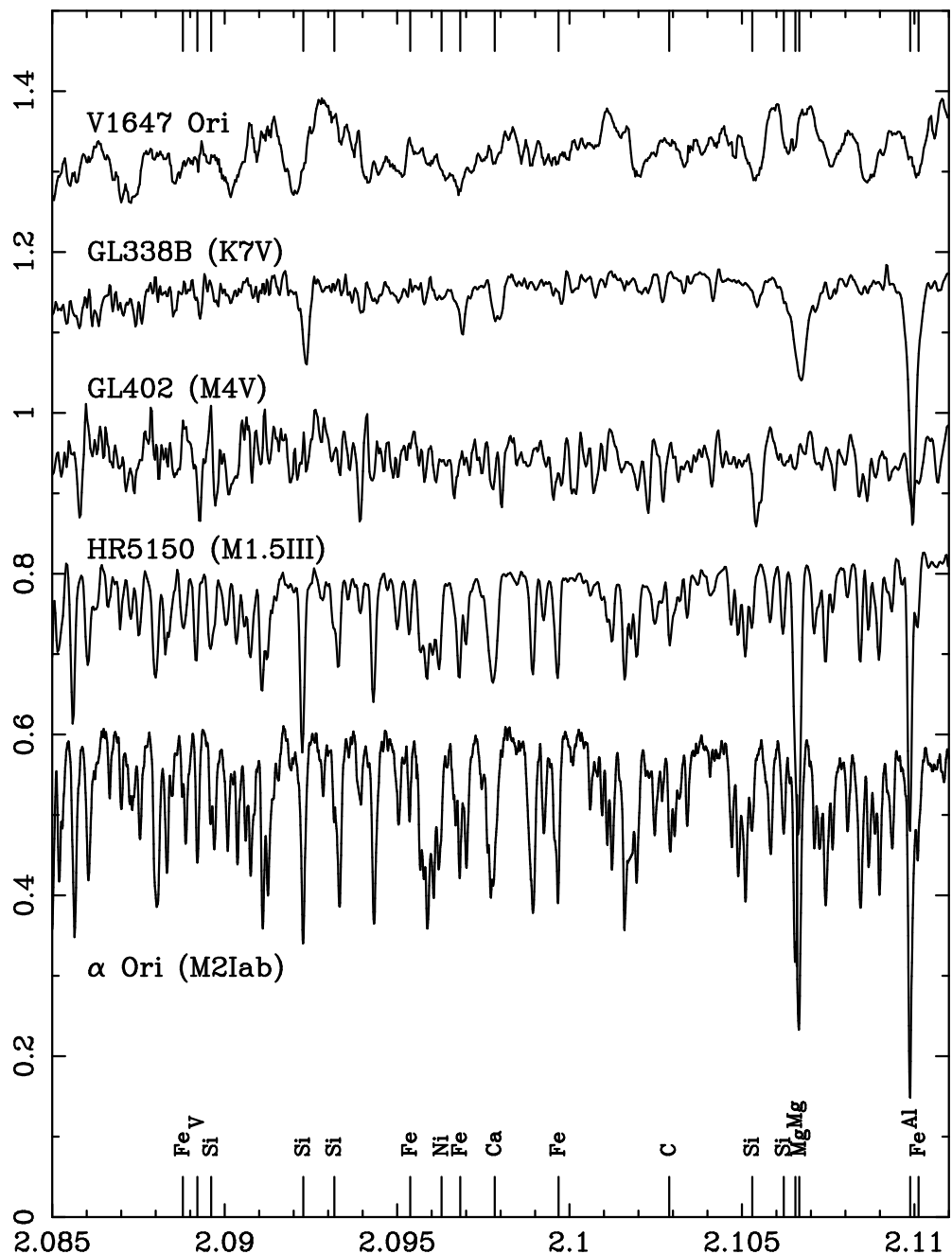} 
\caption{The same plot as in Fig.~\ref{na-mk} but for the echelle order containing Mg~I and Al~I (order \#36).  The spectra have been corrected for the $v_{helio}$ radial velocity at the time of observation.  The same trends are evident as in the Na order, namely that the lines/bands in V1647~Ori are considerably broadened with respect to the corresponding lines/bands in the MK standards stars GL~338B and $\alpha$~Ori.  
\label{mgal-mk}}

\end{figure*}

\begin{figure*}[tb] 
\epsscale{0.8}
\plotone{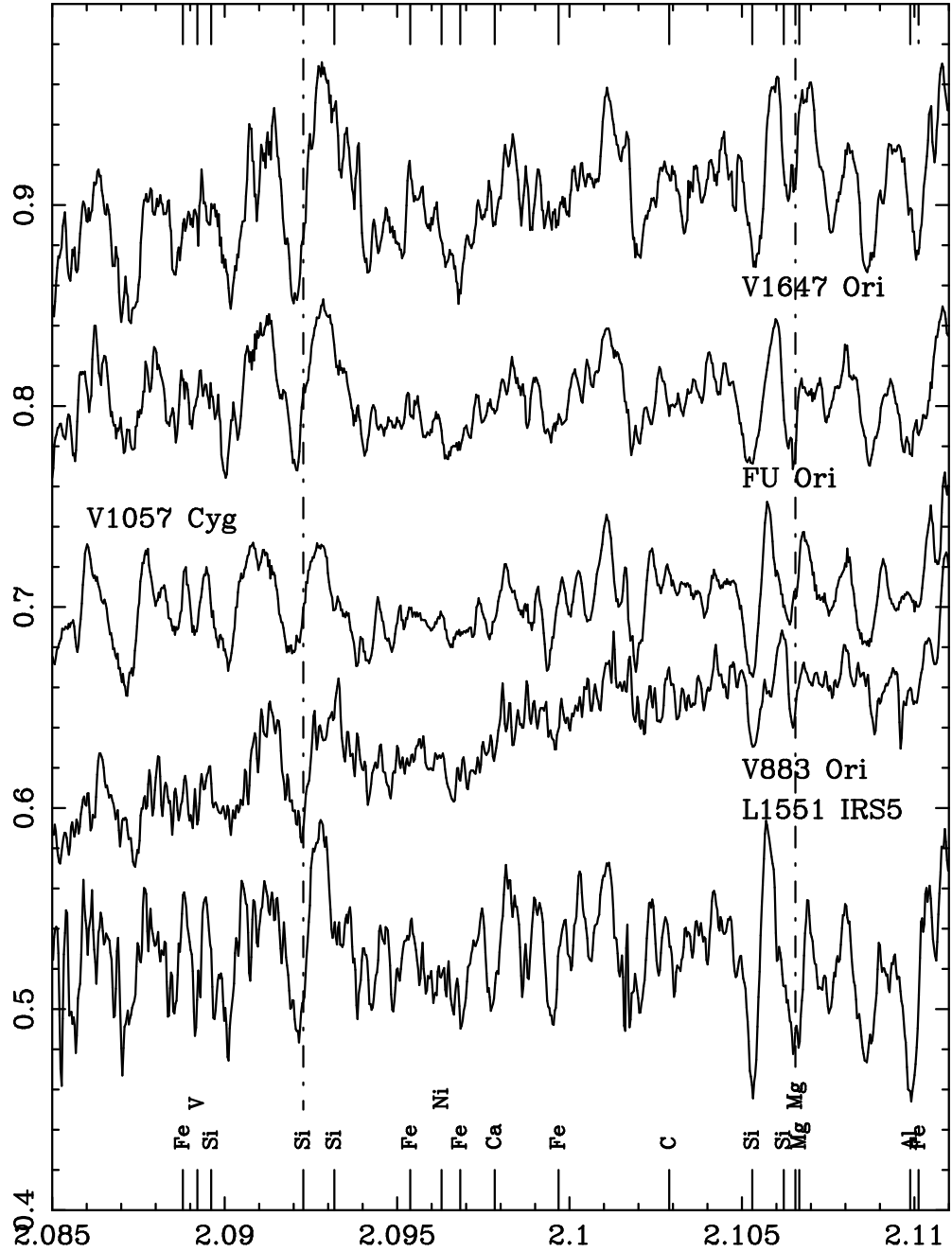} 
\caption{The same plot as in Fig.~\ref{na-fuors-rvcor} but for the Mg~I and Al~I lines in echelle order \#36.  The spectra have been corrected for the $v_{helio}$ radial velocity at the time of observation.  The vertical dot-dashed lines identify two atomic absorption features and are provided to guide the eye between the different spectra.
\label{mgal-fuors-5}}

\end{figure*}

\begin{figure*}[tb] 
\epsscale{0.8}
\plotone{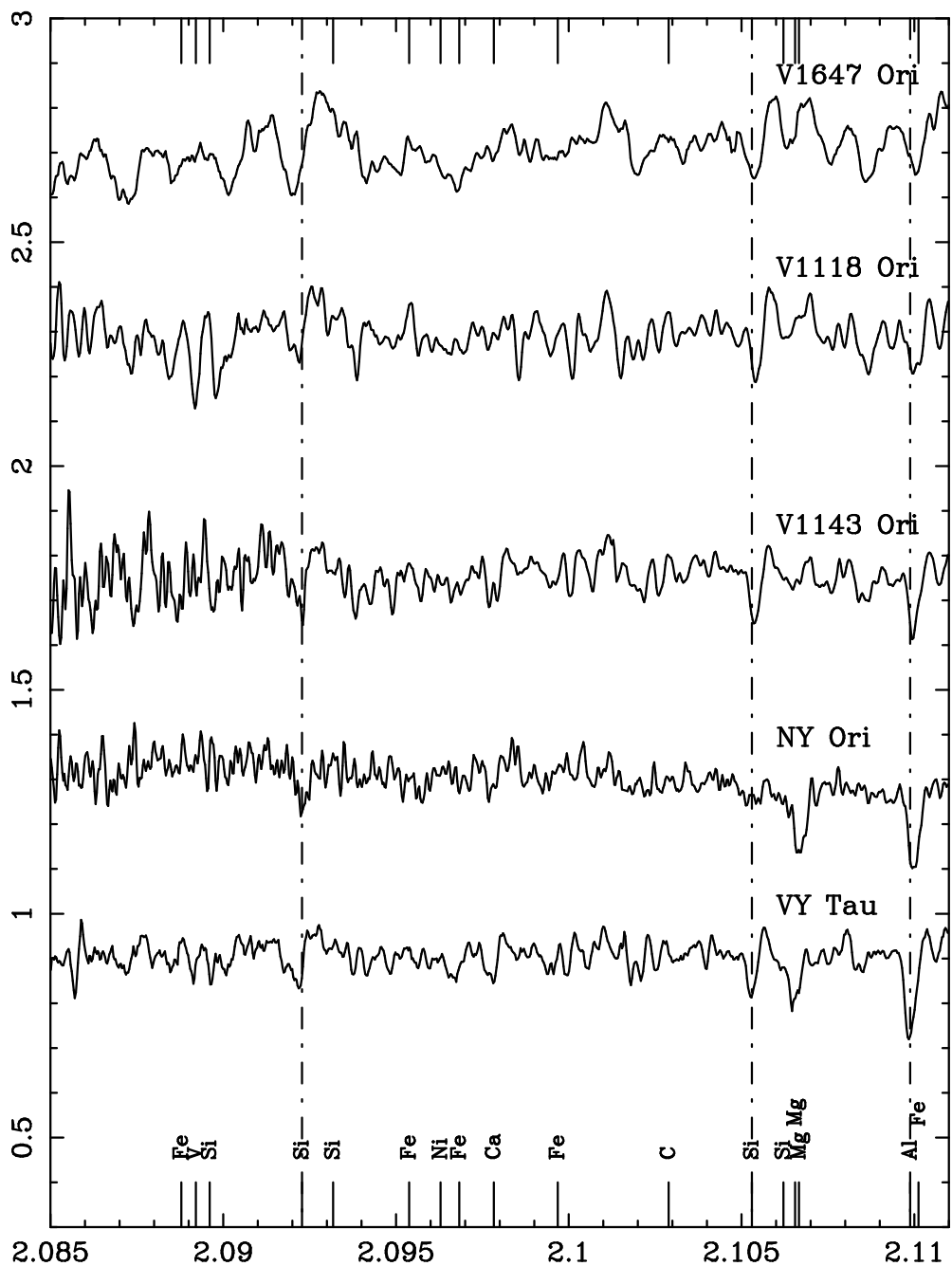} 
\caption{The same plot as in Fig.~\ref{na-exors} but for the Mg and Al lines in echelle order 36.  The spectra have been corrected for the $v_{helio}$ radial velocity at the time of observation.  The vertical dot-dashed lines indicate the location of atomic lines and are provided to guide the eye between spectra.
\label{mgal-exors-full}}

\end{figure*}

\begin{figure*}[tb] 
\epsscale{0.8}
\plotone{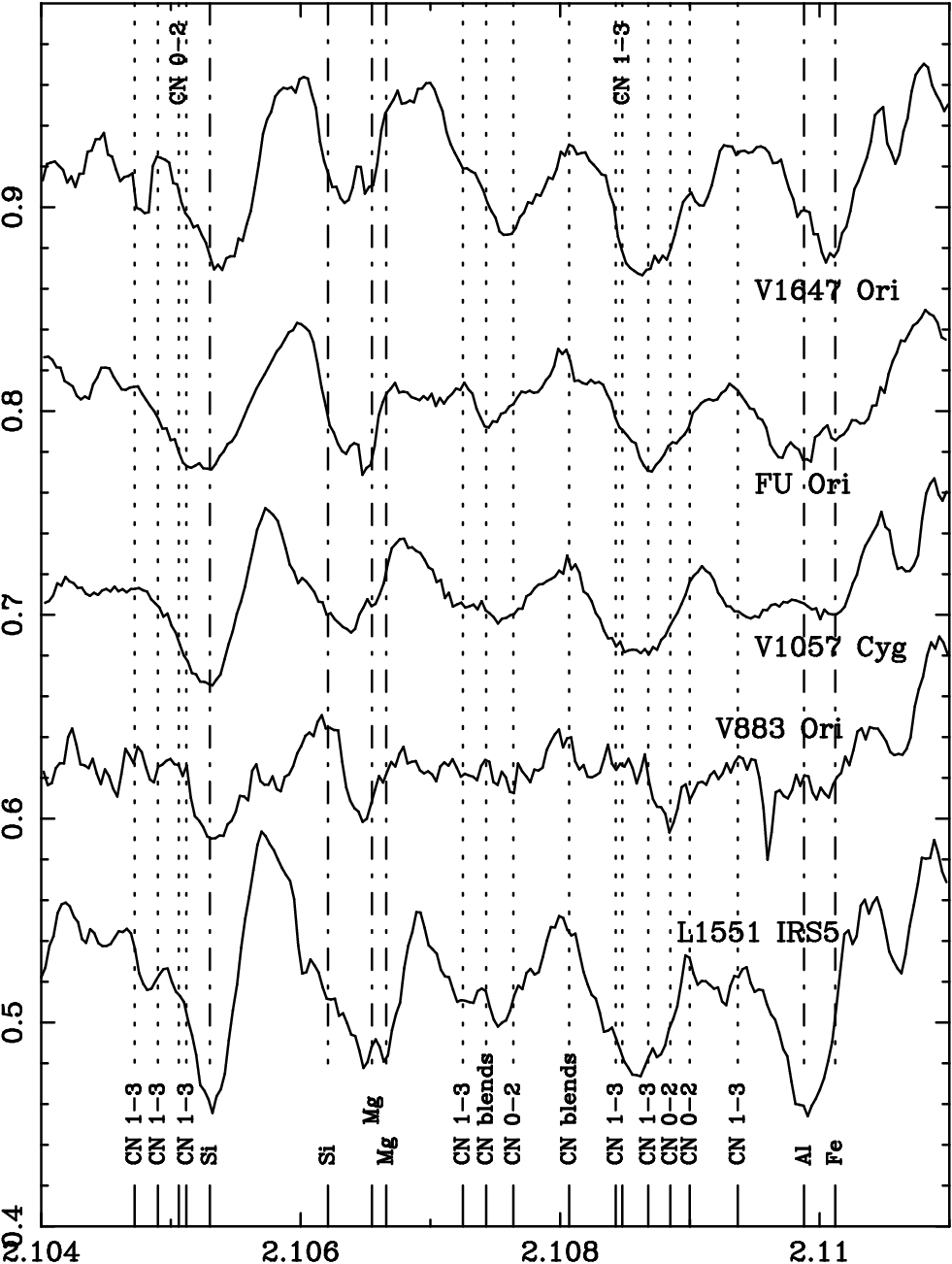} 
\caption{An expanded view of the region around the 2.1065 and 2.1066~$\mu$m Mg~I lines, and the Al~I line at 2.1099~$\mu$m for V1647~Ori and the five FUors shown in Fig.~\ref{mgal-fuors-5}.  The spectra have been corrected for the $v_{helio}$ radial velocity at the time of observation.  The vertical dot-dashed and dotted lines indicate the location of atomic lines and molecular features (specifically CN transitions), respectively.
\label{mgal-fuors-5b}}

\end{figure*}

\begin{figure*}[tb] 
\epsscale{0.8}
\plotone{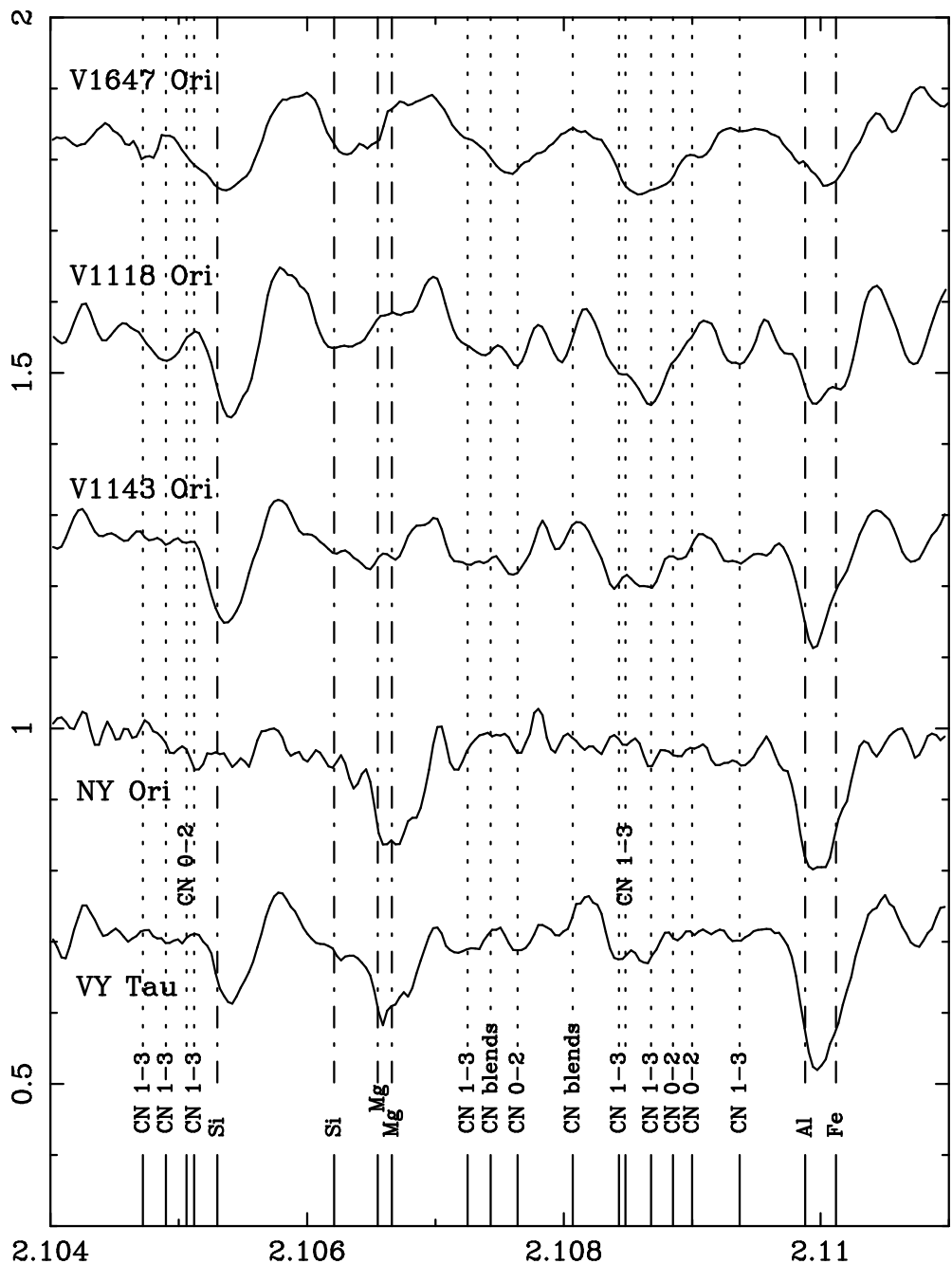} 
\caption{An expanded view of the region around the 2.1065 and 2.1066~$\mu$m Mg~I lines, and the Al~I line at 2.1099~$\mu$m for V1647~Ori and the five EXors shown in Fig.~\ref{mgal-exors-full}.  The spectra have been corrected for the $v_{helio}$ radial velocity at the time of observation.  The vertical dot-dashed and dotted lines indicate the location of atomic lines and molecular features (specifically CN transitions), respectively.
\label{o3-exors}}

\end{figure*}

The structure present in echelle order \#36 of V1647~Ori is very similar in nature to that in the other orders.  The lines are broad and merged and it is difficult to identify discrete features with specific absorption lines.  Fig.~\ref{mgal-mk} shows the V1647~Ori order \#36 spectrum together with the same region from the MK standard stars.  As before, the correspondence in features and line widths is minimal.  A comparison of the V1647~Ori order \#36 spectrum with those of the FUors (Fig.~\ref{mgal-fuors-5}) again shows an excellent agreement in the features present in terms of both location and width.  As before, a comparison with the EXor spectra (Fig.~\ref{mgal-exors-full}) shows a much poorer correlation.  Expanded views of the region around the (temperature sensitive) Mg~I and Al~I lines for FUors and EXors are shown in Figs.~\ref{mgal-fuors-5b} and \ref{o3-exors}, respectively.  Identified on these plots are all atomic (dot-dashed) and molecular CN lines (dotted) present in this region.  Since it is not possible to associate atomic lines with all the absorption features present, it may well be that, for example, the two broad absorption dips between 2.107 and 2.1095~$\mu$m are broadened and merged molecular CN absorption lines.  

The xc of the V1647~Ori order \#36 spectrum with the MK standards gave $<$R$>$=3.3$\pm$3.2.  Only one star produces a statistically significant correlation, namely, GL~402 with R=9.7.  A comparison of the spectrum of GL~402 with V1647~Ori shows that several features present in the M4~V star are also present in V1647~Ori, although they are somewhat broader in the latter.  This correspondence is lacking in a comparison of the other MK standards to V1647~Ori.   For the FUors, the results of the xc with V1647~Ori give R values in a range of 10.5 to 16.9 with $<$R$>$=12.6$\pm$2.1. The same xc with the EXors produces a range of 0.7 to 5.1 with $<$R$>$=3.8$\pm$1.8.  As with the CO order, the correlation coefficient value is over 3$\times$ larger for the FUors than the EXors.  

\subsection{Echelle Order \#35 -- the Br$\gamma$  Line}
In Fig.~\ref{orders4}, echelle order 35 includes Br$\gamma$.  This line is in emission in V1647~Ori and appears considerably broader than unresolved lines with just the instrumental profile.  The FWHM of the Br$\gamma$ line is $\sim$150~km~s$^{-1}$ compared to an unresolved line of FWHM$\sim$17~km~s$^{-1}$.  Due to the Br$\gamma$ absorption in the A0~V telluric standard star, we interpolated across the Br$\gamma$ absorption in the standard to allow the true strength and profile of the Br$\gamma$ line in V1647~Ori (and the other objects) to be studied.  However, due to the spectral resolution of the data and the significant pressure broadening of the Br$\gamma$ in the dwarf atmosphere, we had to interpolate from 2.16 to 2.172~$\mu$m.   This resulted in two strong (and two weak) narrow telluric absorption lines remaining in the V1647~Ori spectrum originating from telluric absorption at $\sim$2.16344~$\mu$m and 2.16869~$\mu$m.  Their presence, however, serves a useful purpose in that they show the width of unresolved spectral features.  In addition to being broadened, the V1647~Ori Br$\gamma$ line profile is a little asymmetric; the long-wavelength side of the profile seems truncated with respect to the short-wavelength side.  This was also seen in H$\alpha$ in the 2007 February 21 optical spectrum presented in ABR08 where it was interpreted as possible evidence for the presence of red-shifted H$\alpha$ absorption from infalling cool gas.  However, due to the presence of weak telluric absorption on the red-wing of the Br$\gamma$ feature, this correspondence cannot be made conclusively.

\begin{figure*}[tb] 
\epsscale{0.8}
\plotone{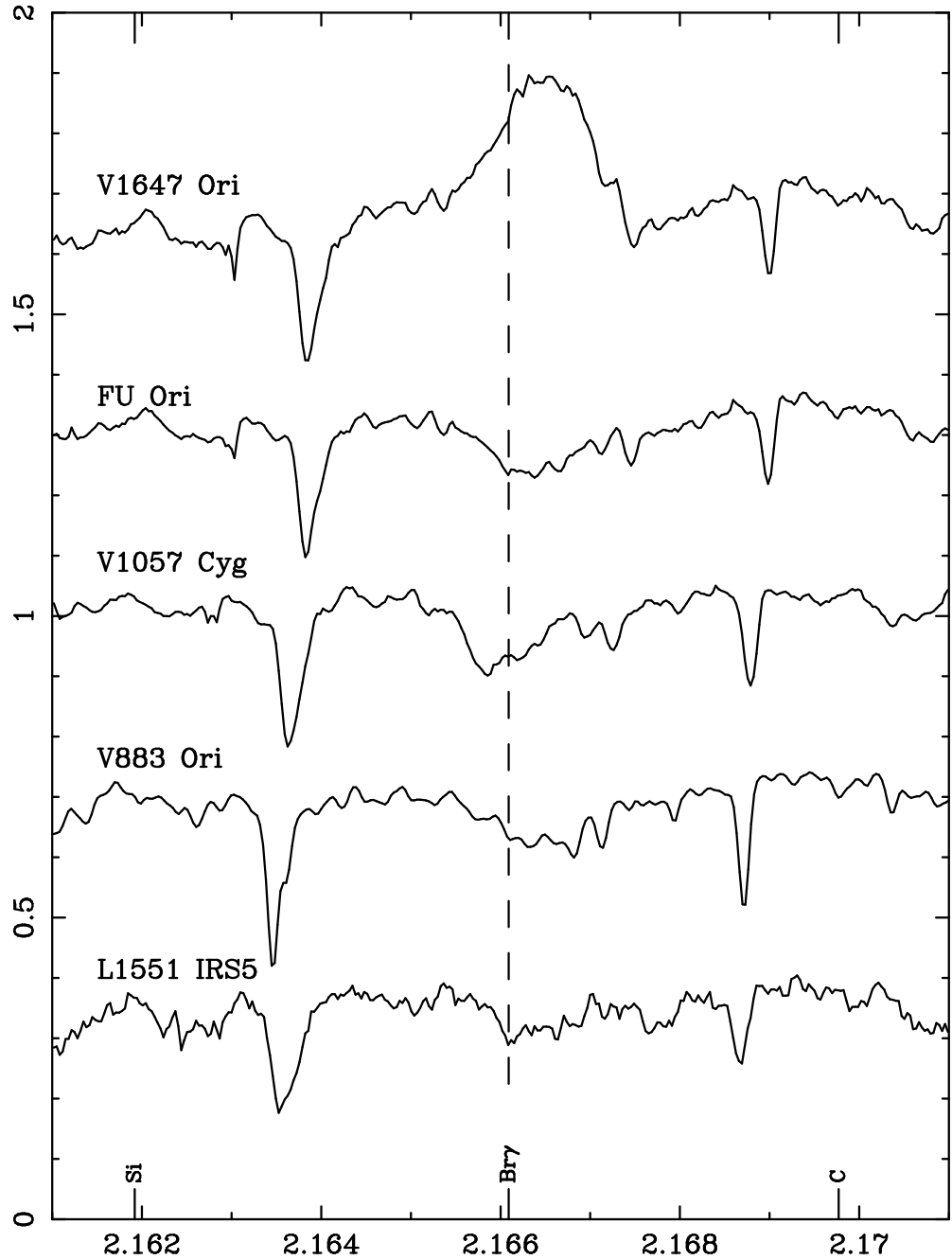} 
\caption{The echelle order containing Br$\gamma$ for V1647~Ori and the four FUors.  Only the region around Br$\gamma$ is shown.  The prominent absorption features shortward and longward of Br$\gamma$ are residual telluric lines due to the interpolation across the Br$\gamma$ absorption feature in the telluric standard.   The spectra have been shifted to correct for {\it v$_{helio}$}.   V1647~Ori shows Br$\gamma$ emission while all the FUors show Br$\gamma$ in absorption. 
\label{o4-fuors}}

\end{figure*}

\begin{figure*}[tb] 
\epsscale{0.8}
\plotone{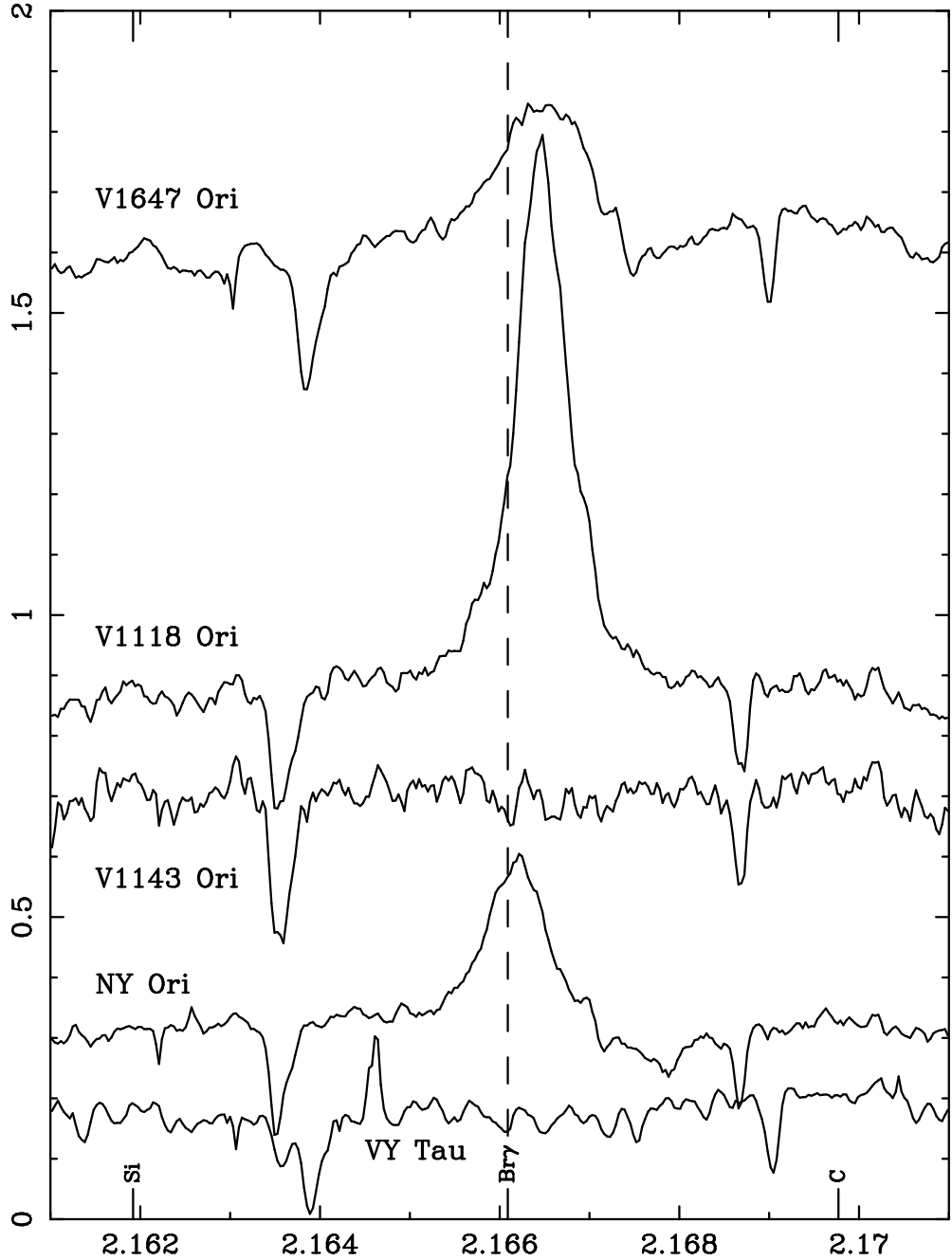} 
\caption{The echelle order containing Br$\gamma$ for V1647~Ori and the four EXors.  Only the region around Br$\gamma$ is shown.  The prominent absorption features shortward and longward of Br$\gamma$ are residual telluric lines due to the interpolation across the Br$\gamma$ absorption feature in the telluric standard.   The spectra have been shifted to correct for {\it v$_{helio}$}.   V1647~Ori and two of the EXors exhibit Br$\gamma$ emission while the other two EXors show no Br$\gamma$ feature. 
\label{o4-exors}}

\end{figure*}

One difference between V1647~Ori and the FUors shown in Fig.~\ref{o4-fuors} is that Br$\gamma$ emission is absent from the FUor spectra.  In fact, our high-resolution spectra of the FUors show, in all cases, weak Br$\gamma$ absorption.  This is not obvious in the lower-resolution spectra of these sources from Reipurth \& Aspin (1997) due to the weakness of the features.  On the other hand, Br$\gamma$ is either in emission or absent in the EXor spectra shown in Fig.~\ref{o4-exors}.  In both V1118~Ori and NY~Ori, Br$\gamma$ has a FWHM of $\sim$100~km~s$^{-1}$.  

\subsection{Repeat Observations of V1647~Ori}
\begin{figure*}[tb] 
\epsscale{0.8}
\plotone{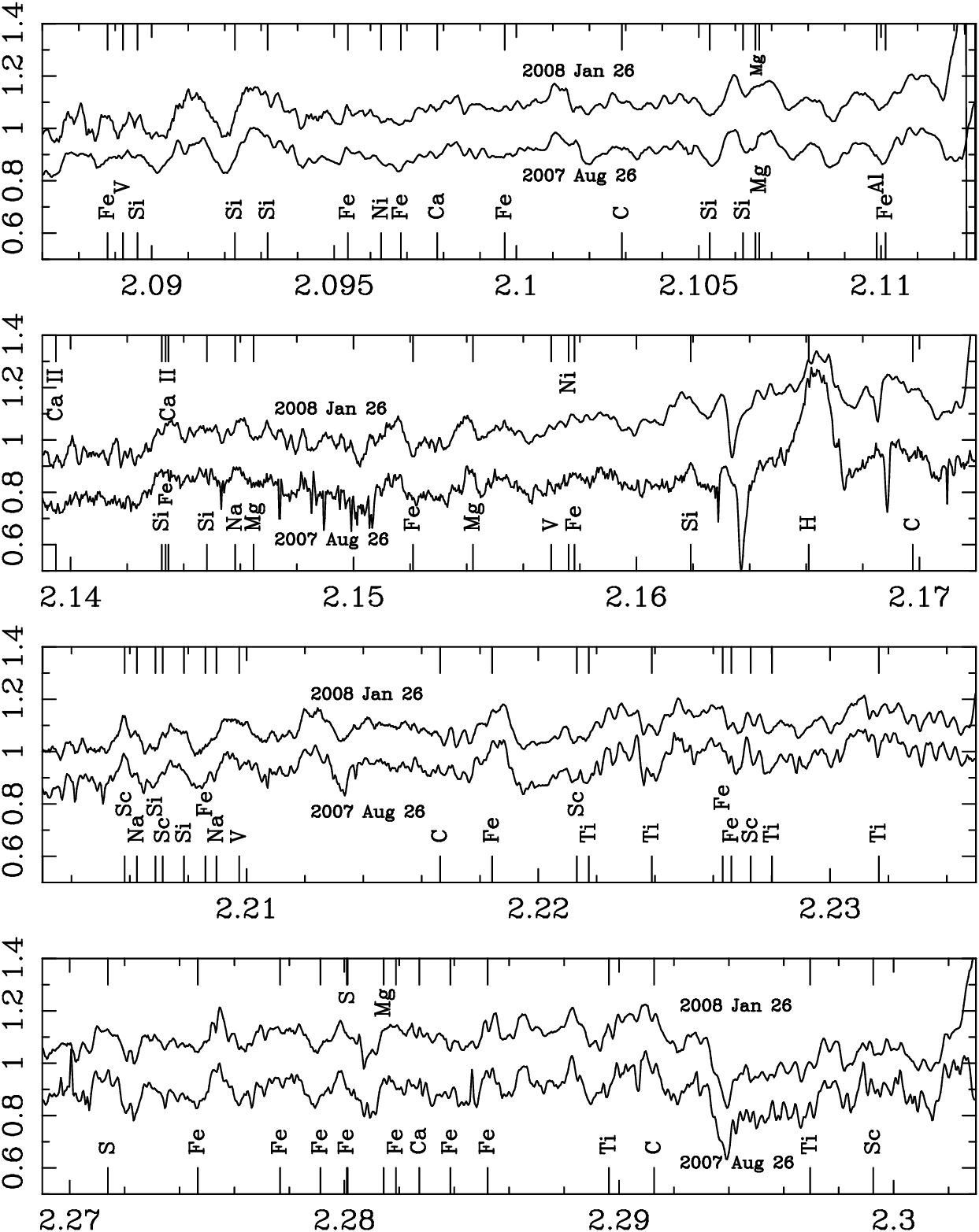} 
\caption{The four echelle orders shown in Fig.~\ref{orders4} for V1647~Ori at two epochs, UT 2007 August 26 and 2008 January 26.  There is good correspondence of features in the two spectra suggesting that the region where these lines and bands are created is relatively stable over periods of several months.  Note that the sharp absorption features at 2.1637 and 2.1688~$\mu$m are residual telluric features and not intrinsic to the object. 
\label{2dates}}

\end{figure*}

We have related above that NIRSPEC echelle observations of V1647~Ori were acquired at two different epochs separated by 6~months.  The same echelle orders as displayed in Fig.~\ref{orders4} are shown in Fig.~\ref{2dates}, this time for both datasets.   The spectral features correlate well in all orders with relatively minor differences.   The Br$\gamma$ emission changed somewhat and weakened between the two observations suggesting that a variation (reduction) in accretion rate occurred.   In a subsequent paper we will investigate the temporal variation in accretion rate estimated from Br$\gamma$ flux since it is beyond the scope of the current work.  The R values obtained from an xc of the two spectra range from 17.3 (\#36) to 21.5 (\#33), strongly supporting the result of the qualitative comparison (see Table~\ref{xcorrnaco} for the complete results).  It seems therefore that, over this 5 month period, the structure seen in the 2~$\mu$m spectrum of V1647~Ori remains relatively stable with the exception of an implied reduction in accretion.  

\subsection{Cross-Correlation of EXors and MK Standards}
In Table~\ref{xcorrnaco2} we also show the results of the xc of the EXor VY~Tau with the other three EXors observed and the MK standards.   The correlation values for all orders in each xc are all greater than the 3.  This implies that the EXors have features in common with each other and, in many cases, many in common with the MK standards.   The highest correlation values found from the xc of VY~Tau and MK standards was for the correlation with the K7~V star GL~338B.  VY~Tau has most in common with the EXor V1143~Ori.

\section{DISCUSSION}
Rather than attempt to definitively categorize V1647~Ori in one of the established classifications, we prefer to simply consider the facts that have become apparent from both the above results and results from other recent work.  From what we have learned we can state that:

\begin{itemize}
\item The high-resolution spectra of V1647~Ori do not resemble those of late-type MK standards (dwarfs, giants, and supergiants) nor a typical Class~I protostar.  This is even the case when the spectra of the MK standards are degraded in resolution to simulate the significant line broadening observed ($\Delta$v$\sim$120~km~s$^{-1}$) in V1647~Ori.  

\item The high resolution NIR spectral properties of V1647~Ori in quiescence do not correlate well with those of known EXors (i.e. NY~Ori, V1118~Ori, V1143~Ori, and VY~Tau).  At lower spectral resolution, EX~Lupi itself showed a dwarf-like K-band spectrum with weak Br$\gamma$ emission during a minor eruption (V$_{max}\sim$11.5) in 1994 (Herbig et al. 2001).  Additionally, during a more significant outburst in 2008 (V$_{max}\sim$8), EX~Lupi exhibited strong K-band emission from the molecular CO overtone bandheads, and atomic Na~I, and Ca~I lines (Aspin et al. 2009).  We note that such an emission spectrum was also observed in V1647~Ori soon after outburst (Reipurth \& Aspin 2004a; Vacca, Cushing, \& Simon 2004) and faded, after several months, to a predominantly absorption spectrum.

\item V1647~Ori shows considerable correspondence with several known classical FUors, a fact supported by the high statistical significance of the xc analyses performed.  We note that, whereas the FUors are mostly in elevated eruptive states during the above observations, V1647~Ori was at its pre-outburst optical brightness, some 18~months after its most recent outburst had subsided.  

\item  ABR08 have shown that the mass accretion rate, one year after the star had supposedly become quiescent, was still considerable at $\sim$10$^{-6}$~M$_{\odot}$~year$^{-1}$.  This suggests that V1647~Ori had not declined to a ``classical'' T~Tauri star (CTTS) state, where the expected accretion rate would be $\sim$10$^{-7~to~-8}$~M$_{\odot}$~year$^{-1}$, but rather, and more appropriately, a Class~I/II state with a typical accretion rate of $\sim$10$^{-6~to~-7}$~M$_{\odot}$~year$^{-1}$.

\item Aspin, Reipurth, \& Herbig (2009), amongst others, have shown that the optical outburst of V1647~Ori, from its initial brightening in late 2003 to its return to a pre-outburst brightness in early 2006, lasted around 27 months.  This timescale is more consistent with an EXor eruption than any known FUor outburst.

\item V1647~Ori had gone into outburst at least one other time, the previous one occurring some 37 years earlier (Aspin et al. 2006).  Again, this is consistent with the behavior of EXors and not that of known FUors.

\item The SED of V1647~Ori pre-outburst, during outburst, and post-outburst, resembled those of known FUors rather than either T~Tauri stars or EXors (\'Abr\'aham et al. 2004; Andrews, Rothberg, \& Simon 2004).

\item The tremendous wind that occurred soon after the outburst, with velocities upwards of 600~km~s$^{-1}$ (Reipurth \& Aspin 2004a), was very reminiscent of those that were seen in V1057~Cyg, a classical FUor, soon after it erupted in 1969 (Herbig 1977, priv. comm.).

\item The majority of FUors show signs of an active molecular CO outflow (Evans et al. 1994; Hartmann \& Kenyon 1996).  This is not the case for V1647~Ori (Lis, Menten, \& Zylka 1999; Andrews, Rothberg, \& Simon 2004).  However, FU~Ori itself shows no evidence of having a molecular outflow either (Bally \& Lada 1983; Evans et al. 1994).  This is also true for the FUor-like close double AR~6A and 6B in NGC~2264 (Aspin \& Reipurth 2003; Moriarty-Schieven, Aspin, \& Davis 2008).

\item It is unclear whether V1647~Ori is associated with Herbig-Haro (HH) objects like many T~Tauri stars and some FUors e.g. V346~Nor/HH57 (Reipurth et al. 1997), PP13S/HH463 (Aspin \& Reipurth 2000), and L1551~IRS5 (Pyo et al. 2008).  Shock-excited optical and NIR emission has been observed in its spectrum (Fedele et al. 2007; ABR08) but it is not yet known if this indicates that a new HH object was ejected during the recent outburst or just represents shocks in the inner regions of the disk.  HH22 and 23 are related to V1647~Ori by proximity, however, it is still to be proven that their origin is V1647~Ori.  It is unlikely that the HH22 flow is driven by V1647~Ori (due to its east-west alignment; Eisl\"offel \& Mundt 1997; ABR08).  However, there is (inconclusive) evidence that HH23 originated from V1647~Ori (ABR08).

\item V1647~Ori is associated with a mm/sub-mm continuum source with a dust mass, M$_{dust}$ of between 0.01 and 0.04~M$_{\odot}$ (Andrews, Rothberg, \& Simon 2004; Tsukagoshi et al. 2005) and Sandell \& Weintraub (2001) have show that known FUors can have dust masses from 0.02~M$_{\odot}$ (FU~Ori) to 0.4~M$_{\odot}$ (V346~Nor).  However, Weintraub, Sandell, \& Duncan (1991) suggests that the disk masses of cTTS's cannot be distinguished from those of FUors.  The conclusion from this is that V1647~Ori cannot be classified on the basis of dust mass but is consistent with many classes of young stars.

\end{itemize}

\section{CONCLUSIONS}
From the data and discussion presented above, it seems clear that V1647~Ori possesses a number of attributes in common with both FUors and EXors.  Its similarity to FUors is highlighted by the striking nature of the structure observed in high resolution NIR spectra, which appears almost identical to that seen in several classical FUors.  Its similarity to EXors is highlighted by its outburst timescale and repetitive nature.  It is perhaps tempting to place this source in a new, intermediate group.  This was recently suggested by K\'osp\'al et al. (2007) and Chochol et al. (2006) where they linked V1647~Ori and the deeply embedded outburst source (DEOS) OO~Ser (Hodapp et al. 1996) as prototypes of a new class of eruptive variables.  However, rather than creating an ad~hoc classification, we prefer to simply conclude that V1647~Ori has characteristics in common with both types of variables.  This perhaps suggests that either the FUor and EXor designations are not as distinct as previously thought (Herbig, private communication), or that FUor events need not span many decades as has been found so far.  It may also hint at an area of commonality in the mechanisms that trigger both EXor and FUor outbursts.  Whichever of the triggering mechanisms discussed in $\S 1$ above is in effect, it is probably going to result in repetitive outbursts (directly observed in EXors). Perhaps some, or all, of the proposed mechanisms may occur, and which one produced the eruption is dependent on the specific details of the physics and geometry of the young star's circumstellar environment and its multiplicity state.

As to V1647~Ori itself, the appearance of its 2~$\mu$m spectrum at R$\sim$~18,000 suggests that the emitting region is influenced by absorption from both atomic and molecular species and this appears to be the case for all FUors observed.  We defer further discussion of the origin and correspondence of high-resolution NIR spectral feature in FUors and EXors to a subsequent paper (Aspin et al. 2009).

We consider that the idea of the existence of a ``continuum'' of eruption characteristics is worthy of further investigation.   The discovery of more eruptive variables and their relationship to the known FUors and EXors is clearly important as is obtaining high-resolution NIR spectroscopy of more EXors especially during outburst and as the event declines.   Such discoveries will be undoubtedly be made through large-scale multi-epoch optical surveys such as those soon to be undertaken by Pan-STARRS-1 (Kaiser et al. 2002) and VYSOS (Reipurth, Chini, \& Lemke 2004).   

\vspace{1.0cm}

The data presented herein were obtained at the W.M. Keck Observatory from telescope time allocated by the University of Hawaii Time Allocation Committee.  The Observatory was made possible by the generous financial support of the W.M. Keck Foundation.  The authors wish to recognize and acknowledge the very significant cultural role and reverence that the summit of Mauna Kea has always had within the indigenous Hawaiian community.  We are most fortunate to have the opportunity to conduct observations from this sacred mountain.  CA acknowledges supported by NASA through the American Astronomical Society's Small Research Grant Program.  TPG acknowledges grant support from the NASA Origins of Solar Systems program WBS 411672.04.01.02 for this work.  BR acknowledges supported from the National Aeronautics and Space Administration through the NASA Astrobiology Institute under Cooperative Agreement No. NNA04CC08A issued through the Office of Space Science.


\clearpage 

\begin{center}
\begin{deluxetable}{lrrrl}
\tablecaption{Journal of Observations\label{jobs}}
\tablewidth{0pc}
\tablehead{
\colhead{Object} & 
\colhead{R.A.} & 
\colhead{Dec.} &
\colhead{Obs. Date} & 
\colhead{S:N\tablenotemark{a}} \\
\colhead{} &
\colhead{J2000} &
\colhead{J2000} &
\colhead{(UT)} &
\colhead{}}

\startdata
L1551~IRS5\tablenotemark{b}        & 04 31 34.1 &  +18 08 05 & 2007~Aug~26 & 100 \\
VY~Tau                             & 04 39 17.4 &  +22 47 53 & 2007~Aug~26 & 85 \\
V1118~Ori                          & 05 34 44.8 & --05 33 42 & 2008~Jan~26 & 115 \\
NY~Ori                             & 05 35 35.8 & --05 12 21 & 2008~Jan~26 & 200 \\
V1143~Ori                          & 05 38 03.9 & --04 16 43 & 2008~Jan~26 & 50 \\
V883~Ori\tablenotemark{b}          & 05 38 18.1 & --07 02 27 & 2007~Mar~06 & 140 \\
FU~Ori\tablenotemark{b}            & 05 45 22.4 &  +09 04 12 & 2007~Mar~06 & 120 \\
V1647~Ori                          & 05 46 13.1 & --00 06 05 & 2007~Aug~26 & 200 \\
V1647~Ori                          & 05 46 13.1 & --00 06 05 & 2008~Jan~26 & 300 \\
$\rho$~Oph~IRS~63\tablenotemark{b} & 16 31 35.5 & --24 01 28 & 2000~May~30 & 140 \\
Par~21\tablenotemark{b}            & 19 29 00.7 &  +09 38 39 & 2001~Jul~07 & 220 \\
HH381-IRS\tablenotemark{b}         & 20 58 21.4 &  +52 29 27 & 2001~Jul~07 & 190 \\
V1057~Cyg                          & 20 58 53.7 &  +44 15 29 & 2007~Aug~26 & 350 \\
HH354~IRS\tablenotemark{b}         & 22 06 50.7 &  +59 02 49 & 2001~Jul~07 & 60 \\
\enddata
\tablenotetext{a}{Signal to noise in the final spectrum.}
\tablenotetext{b}{Observed for, and presented in, Greene, Aspin, \& Reipurth (2008).}
\end{deluxetable}
\end{center}
\clearpage

\begin{center}
\begin{deluxetable}{lccccccr}
\tabletypesize{\scriptsize}
\tablecaption{Cross-correlations\tablenotemark{a}(xc) of V1647~Ori with other objects\label{xcorrnaco}}
\tablewidth{0pc}
\tablehead{
\colhead{Template Object} & 
\colhead{R(36)\tablenotemark{b,c}} & 
\colhead{PH(36)\tablenotemark{d}} &
\colhead{R(34)\tablenotemark{b}} & 
\colhead{PH(34)\tablenotemark{d}} &
\colhead{R(33)\tablenotemark{b,e}} & 
\colhead{PH(33)\tablenotemark{d}} &
\colhead{Classification}}

\startdata
\multicolumn{8}{c}{V1647~Ori (UT~20070826) xc MK Standard Stars} \\
$\alpha$~Ori                  & 2.0   & 0.15 & 0.5  & 0.05 & 7.1(2.4)   & 0.66(0.19) & M2~Iab (observed) \\
$\alpha$~Ori\tablenotemark{f} & 1.6   & 0.14 & 0.3  & 0.04 & 8.4(2.6)   & 0.71(0.33) & M2~Iab (model) \\
GL~338B                       & 1.8   & 0.15 & 0.6  & 0.07 & 9.6(3.8)   & 0.56(0.35) & K7~V \\
GL~402                        & 9.7   & 0.40 & 1.1  & 0.14 & 12.4(4.1)  & 0.64(0.30) & M4~V \\
HR~5150                       & 1.3   & 0.09 & 0.6  & 0.05 & 9.6(3.4)   & 0.59(0.21) & M1.5~III \\
\multicolumn{8}{c}{V1647~Ori (UT~20070826) xc FUors and FUor-like stars} \\
FU~Ori                        & 16.9  & 0.79 & 12.7 & 0.67 & 17.1(13.2) & 0.84(0.79) & FUor \\
HH354~IRS                     & 10.8  & 0.58 & 8.1  & 0.57 & 12.4(10.2) & 0.63(0.48) & FUor-like \\
HH381~IRS                     & 13.3  & 0.78 & 20.9 & 0.80 & 11.8(10.8) & 0.70(0.62) & FUor-like \\
L1551~IRS5\tablenotemark{g}   & 10.5  & 0.59 & 5.6  & 0.50 & 17.7(10.8) & 0.78(0.56) & FUor \\
Par~21                        & 11.9  & 0.72 & 16.4 & 0.71 & 16.2(15.6) & 0.71(0.66) & FUor-like \\
V883~Ori                      & 14.0  & 0.64 & 7.6  & 0.67 & 20.9(13.9) & 0.77(0.69) & FUor \\
V1057~Cyg                     & 11.0  & 0.73 & 7.8  & 0.71 & 17.6(15.7) & 0.77(0.68) & FUor \\
\multicolumn{8}{c}{V1647~Ori (UT~20070826) xc EXors} \\
NY~Ori                        & 0.7   & 0.08 & 0.8  & 0.07 & 4.9(4.1)   & 0.29(0.20) & EXor \\
VY~Tau                        & 5.1   & 0.43 & 1.4  & 0.23 & 11.9(3.7)  & 0.60(0.29) & EXor \\
V1118~Ori                     & 4.8   & 0.44 & 3.9  & 0.29 & 5.8(3.0)   & 0.40(0.18) & EXor \\
V1143~Ori                     & 4.5   & 0.35 & 1.1  & 0.14 & 14.3(5.7)  & 0.63(0.32) & EXor \\
\multicolumn{8}{c}{V1647~Ori (UT~20070826) xc Others Objects} \\
IRS63                         & 1.9   & 0.17 & 0.6  & 0.08 & 13.6(5.0)  & 0.66(0.40) & Class~I protostar \\
\multicolumn{8}{c}{V1647~Ori (UT~20070826) xc V1647~Ori (UT~20080126)} \\
V1647~Ori                     & 16.3  & 0.71 & 19.7 & 0.87 & 18.8(23.5) & 0.86(0.77) & Second date \\
\enddata
\tablenotetext{a}{Calculated using IRAF {\tt rv} package: program {\tt fxcor}.}
\tablenotetext{b}{Tonry \& Davis (1979) R value. R$>$3 implies significant correlation.}
\tablenotetext{c}{R-value calculated over whole echelle order (2.08--2.115~$\mu$m).}
\tablenotetext{d}{Fractional peak correlation function height, in range 0--1.0.}
\tablenotetext{e}{R-value calculated over whole echelle order (2.268--2.305~$\mu$m) and, in parentheses, over the wavelength range 2.27--2.292~$\mu$m (excluding the CO bandhead).}
\tablenotetext{f}{The $\alpha$~Ori spectrum has been veiled by r$_K$=1.0 and broadened by $\Delta$v=120~km~s$^{-1}$.}
\tablenotetext{g}{The 2.122~$\mu$m H$_2$ line was removed from the L1551~IRS5 Na order prior to the xc analysis.}
\end{deluxetable}
\end{center}
\clearpage

\begin{center}
\begin{deluxetable}{lccccccr}
\tabletypesize{\scriptsize}
\tablecaption{Cross-correlations\tablenotemark{a}(xc) of VY~Tau with other objects\label{xcorrnaco2}}
\tablewidth{0pc}
\tablehead{
\colhead{Template Object} & 
\colhead{R(36)\tablenotemark{b,c}} & 
\colhead{PH(36)\tablenotemark{d}} &
\colhead{R(34)\tablenotemark{b}} & 
\colhead{PH(34)\tablenotemark{d}} &
\colhead{R(33)\tablenotemark{b,e}} & 
\colhead{PH(33)\tablenotemark{d}} &
\colhead{Classification}}

\startdata
$\alpha$~Ori                  & 4.6  & 0.30 & 10.2 & 0.53 & 10.7(7.0)  & 0.82(0.43) & M2~Iab (observed) \\
GL~388B                       & 14.3 & 0.68 & 22.3 & 0.87 & 10.0(10.2) & 0.63(0.58) & K7~V \\
GL~402                        & 13.2 & 0.65 & 14.9 & 0.83 & 19.4(7.3)  & 0.81(0.39) & M4~V \\
HR~5150                       & 5.5  & 0.37 & 10.3 & 0.59 & 10.3(7.5)  & 0.66(0.43) & M1.5~III \\
NY~Ori                        & 4.5  & 0.40 & 8.2  & 0.42 & 4.0(5.5)   & 0.41(0.43) & EXor \\
V1118~Ori                     & 5.5  & 0.53 & 6.3  & 0.42 & 5.0(4.6)   & 0.51(0.47) & Exor \\
V1143~Ori                     & 7.6  & 0.58 & 10.7 & 0.76 & 15.3(5.0)  & 0.78(0.46) & EXor \\
\enddata
\tablenotetext{a}{Calculated using IRAF {\tt rv} package: program {\tt fxcor}.}
\tablenotetext{b}{Tonry \& Davis (1979) R value. R$>$3 implies significant correlation.}
\tablenotetext{c}{R-value calculated over whole wavelength range and, in parentheses, the wavelength range 2.104 to 2.111~$\mu$m including just the Mg~I and Al~I lines.}
\tablenotetext{d}{Fractional peak correlation function height, in range 0--1.0.}
\tablenotetext{e}{R-value calculated over whole wavelength range including the CO bandhead and, in parentheses, only the wavelength range 2.27 to 2.292~$\mu$m i.e. excluding the CO bandhead.}
\end{deluxetable}
\end{center}
\clearpage

\clearpage

\clearpage

\clearpage

\clearpage

\clearpage

\clearpage

\clearpage

\clearpage

\clearpage

\clearpage

\clearpage

\clearpage

\clearpage

\clearpage

\clearpage

\clearpage

\clearpage


\end{document}